\documentclass[12pt,letterpaper]{article}

\usepackage[includeheadfoot,
            marginratio={1:1,2:3}, 
            width=412pt,
            height=688pt,]{geometry}
            \usepackage{tikz}
\usepackage[compat=1.1.0]{tikz-feynman}
\usepackage[utf8]{inputenc}
\usepackage{amsmath}
\usepackage{amsfonts}
\usepackage{amssymb}
\usepackage{graphicx}
\usepackage{booktabs}
\usepackage{empheq}
\usepackage{paralist}
\usepackage{cite}
\usepackage[hidelinks]{hyperref}
\usepackage{xcolor}
\usepackage{tensor}
\usepackage{adjustbox}
\usepackage{subcaption}
\usepackage{braket}
\usepackage{yfonts}
\usepackage{slashed}
\usepackage{comment}
\usetikzlibrary{decorations.pathreplacing,calc}
\newcommand{\nc}{\newcommand}
\nc{\lb}{\llbracket}
\nc{\rb}{\rrbracket}
\nc{\gl}{\llbracket}
\nc{\gr}{\rrbracket}
\nc{\bbR}{\mathbb{R}}
\nc{\bbC}{\mathbb{C}}
\nc{\bbZ}{\mathbb{Z}}
\nc{\cO}{\mathcal{O}}
\nc{\cS}{\mathcal{S}}
\nc{\cM}{\mathcal{M}}
\nc{\cT}{\mathcal{T}}
\nc{\cX}{\mathcal{X}}
\nc{\cQ}{\mathcal{Q}}
\nc{\cD}{\mathcal{D}}
\nc{\cC}{\mathcal{C}}
\nc{\cF}{\mathcal{F}}
\nc{\cL}{\mathcal{L}}
\newcommand\beq{\begin{equation}}
\newcommand\eeq{\end{equation}}
\nc{\del}{\partial}
\nc{\tri}{\hspace{-3.5pt}\vartriangle\hspace{-3.5pt}}
\nc{\blacktri}{\blacktriangle}
\nc{\eq}[1]{\begin{equation}
                     \begin{split} #1 \end{split}
                     \end{equation}}
\nc{\ul}{\underline}
\nc{\ov}{\overline}
\nc{\fa}{\hat}
\nc{\fb}{\MakeUppercase}
\nc{\fc}{\tilde }
\nc{\Lie}{{\cal L}} 
\nc{\Tr}{\mbox{Tr}\,} 
\nc{\lambdabar}{{\mkern0.75mu\mathchar '26\mkern -9.75mu\lambda}}

\newmuskip\pFqmuskip

\newcommand*\pFq[7][8]{
  \begingroup % only local assignments
  \pFqmuskip=#1mu\relax
  \mathchardef\normalcomma=\mathcode`,
  \mathcode`\,=\string"8000
  \begingroup\lccode`\~=`\,
  \lowercase{\endgroup\let~}\pFqcomma
  {}_{#2}{#3}_{#4}{\left[\left.\genfrac..{0pt}{}{#5}{#6}\right|#7\right]}
  \endgroup
}
\newcommand{\pFqcomma}{{\normalcomma}\mskip\pFqmuskip}

\allowdisplaybreaks[2]

\makeatletter
 \newcommand{\linkdest}[1]{\Hy@raisedlink{\hypertarget{#1}{}}}
\makeatother

\tikzfeynmanset{warn luatex=false}
\begin{document}

\vspace*{1.5cm}
\begin{center}
{\huge
The Tameness of Quantum Field Theory \\[0.1in]
\large
Part II -- Structures \& CFTs
}
\vspace{.6cm}
\end{center}

\vspace{0.35cm}
\begin{center}
 Michael R.~Douglas,$^{1,2}$
 Thomas W.~Grimm,$^{3}$ and
 Lorenz Schlechter$^{3}$
\end{center}

\vspace{.5cm}
\begin{center} 
\vspace{0.25cm} 
\emph{$^{1}$Center of Mathematical Science and Applications, Harvard University, USA}\\
\emph{$^{2}$
Department of Physics, YITP and SCGP, Stony Brook University, USA}\\
\emph{$^{3}$
Institute for Theoretical Physics, Utrecht University,
%\\Princetonplein 5, 3584 CC Utrecht, 
The Netherlands } \\
   
\vspace{0.2cm}

\vspace{0.3cm}
\end{center}

\vspace{0.5cm}

%%%%%%%%%%%%%%%%%%%%%%%%%%%%%%%%%%%%%%%%%%%%%%%
%%%%%%%%%%%%%%%%%%%%%%%%%%%%%%%%%%%%%%%%%%%%%%%
%%%%%%%%%%%%%%%%%%%%%%%%%%%%%%%%%%%%%%%%%%%%%%%
%%%%%%%%%%%%%%%%%%%%%%%%%%%%%%%%%%%%%%%%%%%%%%%
%%%%%%%%%%%%%%%%%%%%%%%%%%%%%%%%%%%%%%%%%%%%%%%
%%%%%%%%%%%%%%%%%%%%%%%%%%%%%%%%%%%%%%%%%%%%%%%
%%%%%%%%%%%%%%%%%%%%%%%%%%%%%%%%%%%%%%%%%%%%%%%
%%%%%%%%%%%%%%%%%%%%%%%%%%%%%%%%%%%%%%%%%%%%%%%

\begin{abstract}
\noindent
Tame geometry originated in mathematical logic and implements strong finiteness properties
by defining the notion of tame sets and functions. In part I we argued that observables in a 
wide class of quantum field theories are tame functions and that the tameness of a theory 
relies on its UV definition. The aims of this work are (1) to formalize the connection between 
quantum field theories and logical structures, and (2) to investigate the tameness of 
conformal field theories. To address the first aim, we start from a set of quantum field theories 
and explain how they define a logical structure that is subsequently extended to a second structure 
by adding physical observables. Tameness, or o-minimality, of the two structures is then a well-defined property,
and sharp statements can be made by identifying these with known examples in mathematics. For the second aim 
we quantify our expectations on the tameness of the set of conformal field theories and effective 
theories that can be coupled to quantum gravity. We formulate tameness conjectures about conformal field 
theory observables and propose universal constraints that render spaces of conformal field theories to be  tame sets. We test these conjectures in several  examples and highlight first implications.

\end{abstract}

\clearpage

\tableofcontents

%%%%%%%%%%%%%%%%%%%%%%%%%%%%%%%%%%%%%%%%%%%%%%%
%%%%%%%%%%%%%%%%%%%%%%%%%%%%%%%%%%%%%%%%%%%%%%%
%%%%%%%%%%%%%%%%%%%%%%%%%%%%%%%%%%%%%%%%%%%%%%%
%%%%%%%%%%%%%%%%%%%%%%%%%%%%%%%%%%%%%%%%%%%%%%%
%%%%%%%%%%%%%%%%%%%%%%%%%%%%%%%%%%%%%%%%%%%%%%%
%%%%%%%%%%%%%%%%%%%%%%%%%%%%%%%%%%%%%%%%%%%%%%%
%%%%%%%%%%%%%%%%%%%%%%%%%%%%%%%%%%%%%%%%%%%%%%%
%%%%%%%%%%%%%%%%%%%%%%%%%%%%%%%%%%%%%%%%%%%%%%%

\newpage

\parskip=.2cm
\section{Introduction}

Tameness is a concept from mathematical logic, which 
implements a generalized finiteness principle at the most fundamental level. 
In the past years it has been the basis for many remarkable mathematical advances, proving longstanding conjectures,~e.g., in~\cite{pila2014ax,BT,BaldiKlinglerUllmo,AndreOortProof,BTnew}, and giving new proofs to foundational theorems, see e.g.~\cite{zannier2008rational,bakker2020tame}. Even more recently, tameness has found its first applications in physics \cite{Bakker:2021uqw,Grimm:2021vpn,Grimm:2022sbl,Douglas:2022ynw}. In  
ref.~\cite{Bakker:2021uqw}, Bakker, Schnell, Tsimerman and the second author used it to prove a long-standing conjecture of 
Acharya and the first author \cite{douglas_statistics_2003,acharya_finite_2006}, that the number of %supersymmetric
flux vacua in %a GKP type IIB 
certain string compactifications \cite{giddings_hierarchies_2001} is finite. 
In part I of this series \cite{Douglas:2022ynw}, we showed that very general quantum field theories perturbative amplitudes are tame, as are many exact quantum field theory results. 
In this work we aim to significantly extend this perspective and further elaborate on the idea that tameness is a common property of observable physics. We will explain and formalize a new connection between mathematical logic, geometry, and fundamental physics and suggest many applications and future directions. 

In mathematics tameness is implemented by first introducing the notion of a structure and then requiring it to be o-minimal. Such structures can be formulated over the real numbers and collect subsets of all $\bbR^n$s with certain well-defined properties. In mathematical logic, a structure provides a first order language to formulate statements about the real numbers, as we will explain in more detail below. In this context, o-minimality was originally introduced as a means to extend Tarski's decidability theorem \cite{Tarski} to more complicated settings. Remarkably, interpreted geometrically o-minimal structures realize a 
tame topology of the type envisioned by Grothendieck \cite{Esquisse}. On the most basic level, they define what is meant by  a tame set and tame function, which subsequently can be used to define the notion of having a tame geometry.  

There are good reasons to expect that many physical theories will be tame, meaning that their observables, such as correlation functions, partition functions, and amplitudes, are tame
functions.  One reason is that this generalizes the fact that observables are usually real analytic functions
except at phase transitions.
In most examples of o-minimal structures (in fact, in all structures known before 2000) all tame functions can always be decomposed into finitely many 
analytic pieces. % on open sets. 
%such as $\bbR_{\rm an,exp}$ which we review below),
%a function which is piecewise analytic in a compact region is tame.  
The existence of a finite decomposition can be interpreted as having a theory with finitely many phase transitions, a very reasonable requirement.  Tameness can also include
particular non-analytic behaviors such as the essential singularities of the real exponential function at plus and minus infinity.
This behavior is also very common -- it is the $g\rightarrow 0$ limit of the familiar $e^{-1/g^2}$ of instanton and other
tunneling amplitudes, and of the dynamically generated mass scale in an asymptotically free theory.
Thus the intuition that quantum theories become simple at weak coupling can also be expressed by tameness.
Note, however, that one can construct counterexamples, for example by using non-tame functions in the Lagrangian defining
the theory as we have already seen in part I \cite{Douglas:2022ynw}.  Determining precise conditions on a quantum field theory which are sufficient for tameness will be a main goal of the present work.

Another intuition leading to tameness is expressed by the many conjectures that natural sets of theories, such as those
arising from string/M theory compactification or which can be coupled to quantum gravity, are finite
\cite{douglas_statistics_2003,vafa_string_2005,acharya_finite_2006,douglas_spaces_2010,heckman_fine_2019,Hamada:2021yxy}.
Tameness of a set %(very loosely) means 
implies that it is composed of finitely many connected components, so reformulating the
conjectures in this way unifies the case of moduli spaces with that of isolated vacua (without moduli).
In Ref. \cite{Grimm:2021vpn} the second author conjectured that only tame effective field theories can be coupled to quantum gravity.
This suggests that the two intuitions are related, and that the quantum field theories satisfying the conditions for tame observables will be
finite in number. Evaluating this hypothesis will be another main goal of this work. 

Following this discussion, it is natural to split tameness questions in physics into two interrelated parts. We can inquire about the tameness of physical observables of a given, sufficiently constrained, set of theories, and we can ask about the tameness of a set of theories determined by imposing natural physical conditions. As stressed in part I and in \cite{Grimm:2021vpn}, we expect that tameness relies on the UV properties of a theory. For effective theories, we expect that a fixed finite energy cutoff and the compatibility with quantum gravity ensures tameness both for the set of theories and observables. With our current understanding, proving such claims seems far out of reach, but they can still be falsified in specific examples. In contrast, one can hope to eventually prove tameness properties of conformal field theories (CFTs).  
Such theories do not have a distinguished UV scale and we expect the tameness of theory spaces and observables
under a natural set of constraints. In fact, we claim that one can systematically construct o-minimal structures $\bbR_{\text{CFT}d}$ associated to sets of $d$-dimensional conformal field theories. 
The precise constraints that need to be imposed in these constructions will be formulated in two conjectures and it will be a major goal to motivate these by addressing the arising challenges and analyzing explicit CFT examples.

\subsection{Summary of results} \label{summary}

In this section we will give a short summary of the results, leaving most of the technical details for the later sections. Central to this work is the notion of a structure, which can be either viewed as a concept from first order logic or from set theory. The tameness principle is implemented on such structures by requiring them to be o-minimal. There are many known examples of o-minimal structures
and we will introduce several important ones. An introduction to the subject from a logic perspective can be found in section~\ref{sec:structurelogic}, while part I \cite{Douglas:2022ynw} focuses largely on the geometry perspective. 

We will define two structures denoted by $\bbR^{\rm def}_{\cT,\cS}$ and $\bbR_{\cT,\cS}$. The first structure $\bbR^{\rm def}_{\cT,\cS}$ is the structure needed to describe a space $\cT$ of quantum field theories (QFTs) which are well defined in a set of spacetimes $\cS$. Thus it consists of sets of consistent QFTs and the functions required to parameterize these sets.  It summarizes the parameterization of spaces of QFTs (say through couplings in the Lagrangian),  consistency conditions such as unitarity, causality or the absence of tachyons, and it also allows for imposing constraints that restrict to subsets of the theory space, such as a bound on the number of degrees of freedom.

%We will be working mostly with two different structures denoted by $\bbR^{\rm def}_{\cT,\cS}$ and $\bbR_{\cT,\cS}$. The first structure $\bbR^{\rm def}_{\cT,\cS}$ is the structure needed to formulate a chosen set $\cT$ of quantum field theories (QFTs) formulated in a set of spacetimes $\cS$. This structure consists out of the functions required to formulate the quantum field theories as well as additional constraints. These constraints can consist either of consistency conditions like unitarity, causality or the absence of tachyons or of specifically chosen constraints that restrict to subsets of the whole theory space.

The structure $\bbR_{\cT,\cS}$ is the structure needed to formulate general physical statements about the theories $\cT$ on the spacetimes $\cS$.
More precisely, it is obtained by adding all observables, namely correlation and partition functions, of the theories to $\bbR^{\rm def}_{\cT,\cS}$. These observables are functions of the parameters of $\cT$, the positions of the operators, and the parameters labelling spacetime manifolds. 

Next, we discuss general properties of these structures, especially those which bear on o-minimality. For the spacetimes $\cS$ we mainly choose the topologies $S^n$, $T^n$ and $\bbR^n$ or products thereof with conformally flat metrics. While this formulation assumes a non-dynamical background metric, we comment in section \ref{sec:definable-QG} on the structures appearing in theories of quantum gravity. We expect the structures $\bbR^{\rm def}_{\cT,\cS}$ and $\bbR_{\cT,\cS}$ to be tame structures given simple physical conditions on the set $\cT$. These conditions include an upper bound on the number of local degrees of freedom, e.g. on the central charge in CFTs or on the UV central charge for QFTs. Finally, we state precise conjectures for sets of QFTs $\cT$ which, we will argue, have a finite number of components and thus $\bbR^{\rm def}_{\cT,\cS}$ could be o-minimal.  These are
\begin{itemize}
    \item \hyperlink{Conjecture2}{Conjecture 2:} $\cT_{\rm EFT}$, the set of effective theories with a cutoff $\Lambda < M_{\rm pl}$ which can be coupled to quantum gravity; % (and with $M_{pl}>\Lambda$).
     \item \hyperlink{Conjecture5}{Conjecture 5 (a):} $\cT_{\rm CFT2}$ with an upper bound on the central charge and a lower bound on nonzero operator dimensions (the spectral gap of the radial Hamiltonian);
     \item \hyperlink{Conjecture5}{Conjecture 5 (b):} $\cT_{{\rm{CFT}}d}$ with an upper bound on $F$ (in $d=3$) or $a$ (in $d=4$), the quantities which decrease under RG flow.  The precise formulation depends on a point discussed in \S \ref{CFT-tameness} regarding the treatment of theories related by gauging a discrete symmetry.
 \end{itemize}
Furthermore, motivated by a mathematical conjecture,   \hyperlink{Conjecture1}{Conjecture 1}, that adding integrals of tame functions preserves tameness, we formulate the following conjectures on $\bbR_{\cT,\cS}$:

\begin{itemize}
    \item \hyperlink{Conjecture3}{Conjecture 3:} The structure $\bbR_{{\rm EFT}\textit{d}}$ built from all correlation functions for the theories in $\cT_{{\rm EFT}d}$ with cut-off $\Lambda$ on tame spacetime manifolds $(\Sigma,g)$ is o-minimal.
    \item  \hyperlink{Conjecture4}{Conjecture 4:} Given a tame set $\cT_{\rm CFT}$ of CFTs and a tame set $\cS$ of Riemannian manifolds, the structure $\bbR_{\cT_{\rm CFT},\cS}$ is o-minimal.
 \end{itemize}
These conjectures are tested in various examples, including finite dimensional quantum mechanics, 3d Chern-Simons theories, free QFTs in flat as well as AdS spacetimes. We also provide general arguments using known properties of CFTs, e.g., the fact that CFT correlators admit an expansion into conformal partial waves. Our study reveals a number of implicit conjectures. For example, we note that the tameness of observables in interacting CFTs is in close relation with the claims that there cannot be parametrically controlled gaps in the space of operator dimensions (see, e.g.~\cite{Lust:2019zwm}).

\section{Tame structures in first order logic}
\label{sec:structurelogic}

In this section we introduce the notion of an o-minimal structure, but with a focus on its definition in first 
order logic \cite{Wilkie-ominimality}. For readers interested in the more geometric 
approach we refer to part I \cite{Douglas:2022ynw} or the some of the introductory books on the subject such as \cite{VdDbook}. 

The introduction to structures, and tameness, from a first order logic point of view  will later help us in thinking about structures associated to QFTs. In \S \ref{sec:structureQFT} we will construct a structure $\bbR_{\cT}$ that allows us to formulate all `physical' statements about a set of QFTs $\cT$. 
We then introduce o-minimal, or tame, structures, and comment briefly on some of their properties.  

\subsection{A structure and its language}

Here we will define the notion of a structure, sometimes also known as a first-order structure or model. 
We denote a structure $\bbR_{\cF}$
and its logical language $L$ in terms of a base field (here $\bbR$), a set of operations and
a set of functions $\cF$,
\beq \label{structure-withF}
   \bbR_{\cF} :=   \langle \bbR; < , + , \cdot,-, \cF  \rangle\ , \qquad L(\cF)\ .
\eeq
In the following we will make the meaning of these objects more precise. To begin, we  
introduce the simpler structure $\bbR_{\rm alg}$, which has $\cF =\emptyset$. We have defined 
this structure already in part I, section 2, but now we take the logic point of view. 
The basic object is the ordered field of real numbers with a number of standard operations
\beq \label{structureoverlineR}
   \bbR_{\rm alg} = \langle \bbR; < , + , \cdot,-,0,1  \rangle\ ,
\eeq
where $0,1$ are the distinguished elements in the field $\bbR$. 
The definition of $\bbR_{\rm alg}$ comes implicitly with the specification of a logical language $L$. The language uses the formal symbols $x_i$, $i=1,2,...$ together with some notion of polynomials $p(x_1,...,x_n)$.\footnote{At first, the coefficients of these polynomials are assumed to be integers. However, we will be interested in a `structure with parameters', where the coefficients can be real numbers.} These polynomials can be used to state $L$-formulas. By definition $L$-formulas have the properties:
\begin{itemize}
    \item[(i)] The expressions
$p(x_1,...,x_n) = 0$ and $p(x_1,...,x_n) > 0$ are $L$-formulas. They are called atomic $L$-formulas. If $\phi$ is such an $L$-formula then $\phi(\bbR_{\rm alg})$ defines a subset of $\bbR^n$. 
\item[(ii)] If $\phi_1,\phi_2$ are $L$-formulas, then $\phi_1\land \phi_2$, $\phi_1 \lor\phi_2$, and $\lnot \phi_1$ are $L$-formulas. This means that $L$-formulas behave well when applying finitely many logical operations `and', `or', and `not'. In terms of sets, these operations correspond to restricting to the intersection, the union, or the complement of the original set(s). 
\item[(iii)] If $\phi$ is an $L$-formula, then $\exists x_n\, \phi$ is also an $L$-formula. In terms of sets $\phi(\bbR_{\rm alg})=A \subset \bbR^n$ one gets the coordinate projection $\exists x_n\, \phi(\bbR_{\rm alg})=\pi_n (A) $$\subset \bbR^{n-1}.$
\end{itemize}
Hence, an $L$-formula is a formal string of symbols that encodes the construction of a set definable in the structure $\bbR_{\rm alg}$ using the language $L$. Clearly, the structure $\bbR_{\rm alg}$ is rather simple, since it is built entirely from polynomials. It might suffice to formulate simple statements about physical problems that are naturally polynomial, but does not allow us to state more sophisticated questions. In particular, as soon as differential equations are involved one quickly leaves the polynomial world. There is no notion of integral or infinite sum in this language. 

In order to formulate more complicated statements one can extend the structure as in \eqref{structure-withF} by adding a set $\cF$ of real functions. Concretely, this means that one can now form expressions using the ring operations $+,\cdot,-$, real numbers, the formal variables $x_1,...,x_n$, and the functions $f_1,...,f_m \in \cF$ viewed as maps of $(x_1,...,x_n)$ to $\bbR$. Any finite composition of these operations is also allowed in building an expression of $L(\cF)$. 
The notion of $L(\cF)$-formula is then 
the straightforward extension of the conditions (i), (ii), (iii).
In particular, an expression of the form
\beq \label{statements_pf}
   p\big(x_1,...,x_n, f_1(x),...,f_m(x)\big) = 0 \ ,
\eeq
is now an $L(\cF)$-formula, {\it i.e.}~definable in $\bbR_{\cF}$. 
Using $\phi$  to denote the expression \eqref{statements_pf}, we denote by  $\phi(\bbR_{\cF})$ the corresponding subset of $\bbR^{n+m}$. 
As indicated in \eqref{structure-withF} one also allows expressions formed using inequalities.\footnote{
In fact, starting with expressions \eqref{statements_pf} and using the logical symbol $\exists$ one 
automatically gets inequalities.  Consider the set $\{x\in\bbR|\exists y\in\bbR:f(x)=y^2\}$.}

To close this discussion, let us introduce some notation. 
We call a set given by an $L(\cF)$-formula, an $\bbR_{\cF}$-definable set. If it is clear which 
structure we are considering, we will call such sets `definable sets'. We also can
define what is meant by a $\bbR_{\cF}$-definable map, by 
requiring that such a map $f$ between  $\bbR_{\cF}$-definable spaces has 
a graph that is  $\bbR_{\cF}$-definable. Also in this case we might reduce this to the short-hand notation, 
`definable map', and keep the structure under consideration implicit.

\subsection{Tameness -- o-minimal structures} \label{sec:o-minimalstructures}

So far we have imposed no additional conditions on the structure $\bbR_{\cF}$. There is, however, a special class 
of structures, called o-minimal structures, that incorporate a generalized finiteness principle. A structure $\bbR_{\cF}$
is o-minimal, if the only $\bbR_\cF$-definable subsets of $\bbR$ are the finite unions of points and intervals.\footnote{The intervals can be open or closed, and bounded or unbounded.} 
This simple condition on the subsets of the real line impacts the whole structure through its closure under linear projection, i.e.~when using the quantifier $\exists$. Any higher-dimensional set, or associated logical expression, can be reduced to $\bbR$ by projection and this image should only land in the finite collection of points and intervals. We will also refer to o-minimal structures as tame structures, and to their definable sets and functions as `tame sets' and `tame functions'. 

O-minimality has many remarkable implications. Many of these build on a very 
central property of tame sets and tame functions,
the cell decomposition theorem. We describe this theorem in appendix \ref{app-cells}. A cell decomposition of $\bbR^n$ is a slicing of $\bbR^n$ into finitely many pieces called cells, by using definable functions that are also continuous. The cell decomposition 
theorem now states that one can always find such a decomposition such that any definable set is a finite union of cells. Furthermore, for any definable function $f:A \rightarrow \bbR$ one can partition $A$ 
into cells such that $f$ is continuous on each cell. In both parts one can replace `continuous' with
being an element of $C^p$. In many (but not all) o-minimal structures, one also has cell decompositions in which the cells are $C^\infty$ or even analytic~\cite{LionSpeissegger}.

As already indicated, it is often convenient to take 
a geometric point of view on structures built from $\bbR$. If the structures under consideration are o-minimal this leads to the notion of a `tame topology' as envisioned by Grothendieck \cite{Esquisse}. 
To construct this topology one starts with an o-minimal structure $\bbR_{\cT}$ and introduces 
the notion of a definable atlas, which consists of finitely many definable sets with definable transition functions. 
The $\bbR_{\cF}$-definable 
topology can be used to define many other geometric objects that are $\bbR_{\cF}$-definable: manifolds, 
vector bundles, etc. There also exists a powerful extension to complex geometry \cite{PS}.

To indicate the richness of the theory of o-minimal structures, let us list a few examples.
\begin{itemize}
\item $\bbR_{\rm exp}$: $\cF =\{\text{exp}\}$, where $\text{exp}:\bbR \rightarrow \bbR$ is the real exponential function. O-minimality was shown in \cite{Wilkie96}.
    \item $\bbR_{\rm an}$: $\cF = \{ \text{all restricted real analytic functions}\}$. Such functions are all restrictions $f|_{[-1,1]^n}$ of functions $f: U\subset \bbR^{n} \rightarrow \bbR$
that are real analytic on an open set $U \supset [-1,1]^n$. To extend $f|_{[-1,1]^n}$ outside the intervals, one can set it simply to zero there. O-minimality follows from \cite{DenefvdDries}.
    \item $\bbR_{\rm an, exp}$: $\cF = \{\text{exp}, \text{all restricted real analytic functions}\}$.  O-minimality was shown in \cite{DriesMacintyreMarkerRanexp}.
       \item $\mathbb{R}_{\rm Pfaff}$: $\cF=\{\text{all Pfaffian functions}\}$. A function is called Pfaffian if it is part of a Pfaffian chain of functions fulfilling the differential equations $\partial_{x_j} f_i(x)=P_{i,j}(x,f_1(x),\cdots,f_i(x))$, where the $P_{i,j}$ are polynomials of a finite degree. The Pfaffian functions include polynomials, the exponential function as well as all elementary functions on suitable domains.
    
    \item $\mathcal{P}(\tilde{\mathbb{R}})$: Starting with an o-minimal structure  $\tilde{\mathbb{R}}$, one can define its Pfaffian closure $\mathcal{P}(\tilde{\mathbb{R}})$ by adding all solutions to the first order differential equations $\partial_{x_i} f(x)=F_i(x,f(x))$, where all $F_i$ are definable functions in $\mathbb{\tilde{R}}$. O-minimality of $\mathcal{P}(\tilde{\mathbb{R}})$ was proved in \cite{9710220}. 
    
     \item $\bbR_{\cC(M)}$: Start with a series of real numbers $1 \leq M_0 \leq M_1 \leq ...$, such that $\sum_{n=0}^\infty M_n/M_{n+1} = \infty$. To obtain $\cF$ one collects all functions with the following property. Consider $f:[-1,1]^n \rightarrow \bbR$ with $f \in C^\infty$, which satisfy $|f_{\alpha_1...\alpha_n}|\leq c^{|\alpha|} M_{|\alpha|}$, where the $f_{\alpha_1...\alpha_n}$ is the $\alpha_i$'s derivative with respect to $x_i$ and $c$
    is a constant independent of $(\alpha_i,x_i)$. 
    $\cF = \cC(M)$ is the union of all such functions with $n\geq 1$ and generates an o-minimal structure \cite{MR1992825}.
    \item $\mathbb{R}_{\mathcal{G},{\text{exp}}}$: This o-minimal structure consists out of Gevrey functions as well as the exponential function. The Gevrey functions $\mathcal{G}(R,\phi,\kappa)$ are given by all holomorphic functions $f:S\rightarrow \mathbb{C}$ defined on the wedge $S$ in the complex plane, $S:=\{z\in \mathbb{C},0<|z|<R,|\arg(z)|<\kappa \phi\}$, for which constants $A$ and $B$ exist such that $|f^{(n)}(z)|/n!<A B^n(n!)^{\kappa}$ and the limit $\lim_{z\rightarrow 0}f^{(n)}(z)$ exists. \cite{Gstructure} This structure is larger than $\mathbb{R_{\text{an,exp}}}$ and allows to define the $\Gamma$-function on the positive real line.

\end{itemize}
We stress that $\bbR_{\rm an, exp}$ is very rich and allows defining many standard functions. In fact, we have shown in part I that all Feynman integrals are definable in this structure. However, we also made the point in part I that there might be exact (nonperturbative) QFT amplitudes that cannot be defined in $\bbR_{\rm an, exp}$.
Thus it may be relevant that there are structures that are different from $\bbR_{\rm an, exp}$ with more exotic properties. For example, the structure $\bbR_{\cC(M)}$ depends on parameters $(M_i)$ and reduces to $\bbR_{\rm an}$ if one sets $M_n = n!$. This parameterized class of structures $\bbR_{\cC(M)}$ 
can be used to show several important facts about o-minimal structures.
In particular, one can show that there exist two parameter sets $(M_i)$, $(N_i)$ such that $\bbR_{\cC(M)}$ and $\bbR_{\cC(N)}$ are o-minimal without the existence of a bigger o-minimal structure that contains $\bbR_{\cC(M)}$ and $\bbR_{\cC(N)}$ \cite{MR1992825}. This shows that there is no largest o-minimal structure. 

As noted above, not every o-minimal structure admits an analytic cell decomposition. However, it is known that the structures $\bbR_{\rm an}$, $\bbR_{\rm exp}$, $\bbR_{\rm an, exp}$ and $\bbR_{\mathcal{G}}$ do admit such a decomposition \cite{FISCHER2008496}. The Pfaffian closure of a structure admitting analytic cell decomposition is compatible with the decomposition. A counterexample is given by the structure $\bbR_{\cC(M)}$ which does not admit analytic cell decomposition \cite{MR1992825}. The
examples that do not admit such a decomposition will play no role in this work.

Let us close this discussion of tameness by noting that there are other logical properties that one might want to impose on a structure. Most notably, one might want to ask about `model completeness' and about `decidability' of $\bbR_{\cF}$. We refer the reader to \cite{Krapp2019} for an introduction to these concepts. Neither of these properties is implied by tameness, but they are known to hold for many o-minimal structures \cite{Wilkie1996ModelCR,MR1992825}.

\subsection{Physical application to QFT} \label{sec:examples}

In the following we give some examples of physically interesting statements which can be made using a structure $\bbR_{\cF}$. 

We first exemplify some aspects of definable Lagrangians $\cL$ of QFTs. We start with the abstract variables $\{x_i\}_{i=1,2,...,m+n}$ used in some logical language $L(\cF)$. In our example, we split them into two sets, $\{x_i\equiv \lambda_i\}_{i=1,...,m}$ and $\{x_{\alpha+m}\equiv \phi_\alpha\}_{\alpha=1,..,n}$. $\lambda_i$ will be coupling constants or parameters in the Lagrangian, and $\phi_\alpha$ scalar fields. Let  $V(\lambda,\phi)$ be the scalar potential (or any other coupling function) which is a function of $\lambda,\phi$.
If $V$ is an algebraic function, {\it i.e.}~defined by solving polynomial equations, it is definable in the structure $\bbR_{\rm alg}$. In fact, it is definable in every structure, since each $\bbR_{\cF}$ is obtained by extending $\bbR_{\rm alg}$. Of course, one might also want to consider more general scalar potentials and coupling functions of $(\lambda,\phi)$. In this case, $V(\lambda,\phi)$ needs to be definable in $\bbR_{\cF}$. This can always be achieved by adding $V$ to the set of functions $\cF$. 
Assuming that $V$ is twice differentiable, we can then formulate in $\bbR_{\cF}$ statements like 
\begin{align} \label{logic-statements}
  &(1)\quad \exists \phi_\alpha: V(\lambda,\phi) > 0\ \land \ \partial_{\phi_\alpha} V(\lambda,\phi) =0\ \land \lambda_\alpha>0\ , \nonumber \\
  &(2)\quad V(\lambda,\phi)=0 \  \lor\  \partial_{\phi_\alpha}\partial_{\phi_\alpha} V(\lambda,\phi) < 0\ , \qquad etc.
  \end{align}
These expressions are composed of finitely many logical operations among equalities and inequalities involving definable functions, where one uses that if $V$ is definable in $\bbR_{\cF}$ then so are its derivatives.\footnote{To see this, one uses that the mathematical definition of the derivative can be read as a logical statement involving a finite number of additional variables ($\epsilon$'s and $\delta$'s).
} Hence, the expressions are $L(\cF)$-formulas, i.e.~definable in $\bbR_{\cF}$.   

We stress that there is no {\it a priori} restriction on the functions we can include in a structure, but we might need to be careful about the details to get an o-minimal structure. For example, if we want to formulate statements about $V(\lambda, \phi) = \lambda \cos(\phi)$ with $\phi \in \bbR$, we should choose a structure containing $\cos(x)$, such as $\bbR_{\cos}$ which has $\cF=\{\cos(x)\}$.
A statement such as $V(\lambda,\pi \phi)=0$ is then definable in the structure $\bbR_{\cos}$ and defines the integers $\bbZ$. Evidently, the structure $\bbR_{\cos}$ cannot be o-minimal.  On the other hand, suppose we constrain the domain of $\phi$ to the finite length interval $[0,a]$,  Then we can instead add a function $f(x)$ to $\cF$ which is $\cos(x)$ on $[0,a]$ and $0$ otherwise. This structure is o-minimal, since it is a substructure of $\bbR_{\rm an}$. 

Let us now ask whether a structure $\bbR_{\cF}$ in which the classical Lagrangian $\cL(\lambda,\phi)$ is definable is expected to be sufficient to make statements about the quantum theory. Generally, we consider $\cL$ to a functional
of fields $\phi$ on a $d$-dimensional spacetime $\Sigma$. Quantum correlation functions for local operators $\cO_i(y_i)$ at some spacetime positions $y_i$ take the form
\beq \label{correlation_fct_x}
 \langle \cO_1(y_1)...\cO_k(y_k) \rangle = \int D\phi \, \cO_1(y_1)...\cO_k(y_k)  \, e^{-\int_\Sigma d^d y \cL }  \ ,
\eeq
{\it i.e.}~are given by a path integral over the fields $\phi$. We can now ask if the correlation functions are definable in $\lambda$ in the original $\bbR_{\cF}$. 

One immediately sees that this is a very non-trivial question, since \eqref{correlation_fct_x} involves non-definable operations and objects. Most importantly, the notion of integrating a function is not definable: it is not guaranteed that the integral of a function definable in $\bbR_{\cF}$ is again definable in the same structure. A simple example of this was given in part I where we showed that a 0-dimensional QFT with a $\phi^4$ Lagrangian, which is clearly definable in $\bbR_{\rm alg}$, leads to a partition function given by the modified Bessel function $K_{1/4}(x)$. This Bessel function is clearly not definable in $\bbR_{\rm alg}$.  We could continue to list other correlation functions in this theory, which also need not be definable in $\bbR_{\rm alg}$.

Now we can explain the main new concept introduced in this paper.
Given a list of correlation functions for some QFT or set of QFTs,
we can add them to our function set $\cF$ to get
what in general will be a new structure, which characterizes the QFT or QFTs. We define this procedure next.

\section{Structures from physical theories} \label{sec:structureQFT}

In this section we discuss how one can systematically associate two structures $\bbR^{\rm def}_{\cT,\cS}$ and $\bbR_{\cT,\cS}$, obeying the axioms defined in \S \ref{sec:structurelogic}, to a given set $\cT$ of physical theories and a set $\cS$ of spacetimes. The basic idea is that the sets $\cT,\cS$ are definable in $\bbR^{\rm def}_{\cT,\cS}$ and all physical questions, i.e.~statements about the observables of the theory, should definable in $\bbR_{\cT,\cS}$. In order to do that we propose which 
functions of the theory need to be added to the set of functions generating the 
structure.  While we could simply add every function that is needed to work with the theories under consideration, the key point is to be selective, so that the considered structures have a chance to be o-minimal. In \S\ref{sec:tameobservables} we will take a closer look at $\bbR_{\cT,\cS}$ and discuss its tameness assuming an o-minimal $\bbR_{\cT,\cS}^{\rm def}$. It will then be the task of \S \ref{Tameness-cT} to propose conditions on $\cT$ such that $\bbR_{\cT,\cS}$ satisfies the tameness criterion. We mainly discuss structures associated to quantum field theories (QFTs)
in \S \ref{QFT-structure}, \S\ref{sec:definable-spacetime} and briefly comment on theories with gravity in~\S \ref{sec:definable-QG}.

\subsection{Structures from QFTs} \label{QFT-structure}

Our first goal is to describe a procedure which, given a set of QFTs parameterized by $\cT$ and a set of spacetimes $\Sigma$ with metrics $g$ labelled by $\cS$, produces two structures $\bbR_{\cT,\cS}^{\rm def}$ and $\bbR_{\cT,\cS}$. 
While the structure $\bbR_{\cT,\cS}$ allows formulating statements about the observables,
$\bbR^{\rm def}_{\cT,\cS}$ only allows formulating statements about the definition of the set of theories and spacetimes.

While we believe that our definitions can be made mathematically rigorous, we will refrain from becoming too technical when describing QFTs on a curved spacetime.  
The starting point for a more complete discussion would be a rigorous definition of QFTs and correlation functions.  Of the vast body of
work in this area let us cite \cite{Jaffe:2006uz,costello2021factorization} as
examples which the reader could keep in mind for the following.

\subsubsection*{Defining a structure}

Let us consider a set of QFTs labelled by a parameter set $\cT$.\footnote{We will sometime call $\cT$ the 
space of QFTs but we implicitly assume that it also captures the information about the QFTs. In mathematical terms, $\cT$ can be thought of as a moduli space whose points are objects with additional data.} 
The details on how to define these theories depends on the framework (operator formalism, path integral, {\it etc.}) and type of theory. As an example, consider a set of QFTs specified by a family of Lagrangians. As in \S\ref{sec:examples} we can require that the Lagrangians $\cL$ are definable in some structure.
The Lagrangians can depend on dynamical fields and on couplings $\lambda$ parameterizing 
specific QFTs in the family.
We then impose constraints on the couplings $\lambda$, which for simplicity can think of a $\lambda \in \bbR^k$, such that the quantization of the corresponding action produces a well-defined and nonsingular QFT, meaning that all of the observables we are about to list are well defined and satisfy our axioms for QFTs. We require that the set $\cT$ maximally consists of all these 
values of $\lambda$, i.e.~view $\cT$ as a subset of $\bbR^k$. As we will see later on, we will often impose additional conditions on the theories in $\cT$, e.g.~a bound on the number of degrees of freedom, in order that $\cT$ can possibly be a tame set. In other words, while there are natural constraints on $\cT$ such as well-definiteness, we are free to reduce this space further. We denote the structure in which the space $\cT$ and the theories themselves can be defined by $\bbR_{\cT,\cS}^{\rm def}$. 

The structure $\bbR_{\cT,\cS}^{\rm def}$ needs to not only define the theories, but also the spacetime background on which we evaluate the them to compute physical quantities. Let us
start with a set $\cS$ of allowed spacetimes $\Sigma$ with metrics $g$ \footnote{Correlation functions are formulated on punctured spacetimes with operator insertions at the punctures. Removing a finite number of points from a tame set preserves the tameness, thus spacetimes with labeled punctures can be defined if the spacetime itself is definable.}.
We can think of these as being labelled by some parameters $\rho$, {\it i.e.}~consider $(\Sigma,g)_{\rho}$ fibered over some parameter space. This parameter space and the spacetimes with metrics should be definable in $\bbR_{\cT,\cS}^{\rm def}$. While the choice of $\Sigma_\rho$ can affect the result,
this is not our primary interest and thus we will often use the notation $\bbR^{\rm def}_\cT$
to mean that $\Sigma_\rho$ is a standard choice, which we will take to be the conformally flat manifolds.

Having described the construction of $\bbR_{\cT,\cS}^{\rm def}$, let us now turn to $\bbR_{\cT,\cS}$. 
We then choose a set of functions in terms of observables $O_{\kappa}$ computed within the QFTs in $\cT$. These are functions on $\cT\times(\Sigma_\rho)^n$, 
\beq
   O_{\kappa}(y_1,...,y_n;\lambda,\rho)\ , \qquad y_k \in \Sigma_\rho\,,\  \lambda \in \cT\ ,\ \rho \in \cS\ , 
\eeq
where $y_k$ are the $n$ spacetime positions. 
We now introduce formal variables $x_i$ parameterizing 
$\bbR^m$, for a sufficiently large $m$. Considering local neighbourhoods around the $y_i \in \Sigma_\rho$, we can then introduce local coordinates on the product that specify 
the vector $(y_1,...,y_n)$. The latter we 
identify with $n\cdot \dim \Sigma_\rho$ variables $x_i$. The remaining variables are the parameters $\lambda,\rho$ and depending on the way we describe $\cT,\cS$, we 
can again pass through some local construction. In this way we understand $O_{\kappa}(x)$ as a function in some region of $\bbR^m$, which we can extend to all of $\bbR^m$ by 
setting it to zero everywhere else. We now  
can define a structure $\bbR_{\cT,\cS}$, which extends $\bbR_{\cT,\cS}^{\rm def}$ with the functions 
\beq 
   \cF = \Big\{ \bigcup O_{\kappa}\Big\}\ ,
\eeq 
where the union is taken over all observables and over all regions on $\Sigma_\rho$ and $\cT$ in which they can be evaluated. Note that $\bbR_{\cT,\cS}$ is the `quantum structure' associated to the sets $\cT,\cS$, since we have built it 
from quantum correlation functions. Even for QFTs, {\it e.g.}~with polynomial Lagrangian, we expect  
that $\bbR_{\cT,\cS}$ contains many non-trivial functions as we have seen in 
part I, \S 4. 

Our requirement that $\cT$ and $\Sigma_\rho$ are $\bbR_{\cT,\cS}^{\rm def}$-definable gives us a systematic way of extracting the information relevant for $\cF$ from the functions $O_{\kappa}$ on $\cT\times(\Sigma_\rho)^n$.
We recall from \S \ref{sec:structurelogic} that a definable manifold can be covered by coordinate
patches which are mapped by definable functions into definable subsets of $\bbR^d$.  We use this to
combine all of the observables defined on any of the $(\Sigma_\rho,g_\rho)$ into a single structure.

The observables we will allow are canonically normalized partition functions
and correlation functions
taking values in $\bbR$
and defined on a family of spacetime manifolds $\Sigma_\rho$. In other words, we consider 
\beq \label{correlators_1}
   O_{\kappa}(x) = \langle\cO_1 (y_1) \ldots \cO_n (y_n) \rangle_{\lambda,\rho}\ , 
\eeq 
where $\{\cO_1,...,\cO_n\}$ is a set of local operators of the QFT with parameters $\lambda$. $\kappa$ is a formal label of this set of operators. 
If a correlation function is complex valued we take its real and imaginary parts. Note that many observables are naturally thought of as taking values in vector bundles over $\Sigma_\rho$, $\cS$ and $\cT$.  Furthermore we need
to say how the operators depend on the QFT parameters.  These points will be discussed further below.

A basic example for \eqref{correlators_1} would be a two-point function in quantum mechanics related to
a transition of energy $E$.  It is a function of the elapsed time $\Delta t$ between the
operators. In real time quantum mechanics it would be $e^{iE|\Delta t|}$, which
is not tame.  Thus (in this work) we take $\Sigma_\rho$ to be Riemannian (not Lorentzian).  
In particular QFT1 is Euclidean time quantum mechanics,
so the normalized two-point function for an operator which induces a
transition of energy $E$ is $e^{-E|\Delta t|}$.

Generally speaking, partition and correlation functions are only defined up to an overall multiplicative factor,
which can depend on the parameters $g\in\cT$ and $\rho\in\cS$.  
By canonically normalized, we mean that there must be some physical condition which
removes this ambiguity.  Usually this will be done by interpreting the observables in the operator
formalism.  For example, a partition function on $\Sigma=S^1\times X$
(with translation on $S^1$ acting by isometries) has a more fundamental
definition as the trace $\Tr \exp (-LH)$ where $L$ is the circumference of $S^1$ and
$H$ is the Hamiltonian for canonical quantization on a hypersurface $X$.
On the other hand the partition function on $S^d$ is not canonically normalized
and will not be interpreted as a definable function.\footnote{
There can be particular universal terms such as a volume independent or log volume
term in an expansion of $\log Z$.  In fact the volume independent term in $d=3$ will 
play a role below, however it will not be declared to be definable.}

Correlation functions can also be canonically normalized by their relation to the
operator formalism.  In conformal field theories (CFTs) this is done by the state-operator correspondence, which
is a bijective map between the space of local operators and the Hilbert space of
radial quantization.  Thus we can use unit normalized operators in correlation
functions and remove the overall ambiguity for a given $(\Sigma,g)$ by dividing by
the correlation function of the identity operator.  In an asymptotically free or
UV safe QFT this could be done by defining the normalizations in the UV limit.\footnote{
In a QFT with an S-matrix one defines normalized on-shell momentum space
correlation functions using the LSZ formalism, as was done in part I.}

\noindent
In summary, the structure $\bbR_{\cT,\cS}$ (or $\bbR_{\cT}$ when we fix a canonical choice for $\Sigma$) is built from the 
following data:
\begin{enumerate}
    \item A structure $\bbR_{\cT,\cS}^{\rm def}$ that allows to define theories with labels in $\cT$ and spacetimes with labels in $\cS$:
    \begin{itemize}
    \item A set $\cS$ labelling $\bbR^{\rm def}_{\cT,\cS}$-definable Euclidean spacetimes $(\Sigma_\rho,g_\rho)$ with $\rho\in\cS$.
    If not explicitly stated, $\cS$ is the set of $d$-dimensional conformally flat
    manifolds (as discussed below).
    \item A set $\cT$ labelling $\bbR_{\cT,\cS}^{\rm def}$-definable, consistent QFTs. $\cT$ might collect discrete or continuous parameters in the Lagrangian,
    or data such as operator dimensions and coefficients of the operator product expansion. The set $\cT$ 
    can be constrained by additional conditions on the considered theories. 
    \end{itemize}
    \item Canonically normalized partition and correlation functions \eqref{correlators_1} which depend on the parameters $\lambda \in\cT$ and $\rho\in\cS$, a list of local operators and
    their positions in the spacetimes $\Sigma_\rho$.
\end{enumerate}
Note that $\cT$ and the correlation functions need {\bf not} be definable in the same structure as the Lagrangians are. For example, one might want to impose a unitarity bound on the theories labeled by $\cT$ and this could require to perform complicated manipulations with the Lagrangian. Consider, for example, the 0d $\phi^4$-theory as in part I, i.e.~a theory with a polynomial action $S=\frac{m_0^2}{2}\phi^2+\frac{\lambda}{4!}\phi^4$. The correlation functions of this model are not in $\bbR_{\rm alg}$. It can now happen that a condition on the quantum corrected parameters, such as the renormalized mass $m=m(m_0,\lambda)$, cannot be formulated in the original structure $\bbR_{\rm alg}$. While it is not immediately obvious that 
%We do not have an explicit example in which 
$\cT$ itself is not $\bbR_{\rm alg}$-definable, a generic situation where this could happen is a parameterized set of theories with a scalar mode
of mass squared $m^2(\lambda)$ which takes both positive and negative values.  Of course the negative
values lead to instability (a ``tachyon'').  This could result in a phase transition, in which case
we expect that correlation functions are discontinuous at $m^2(\lambda)=0$.
Or it could lead to ill-definedness, in which case $m^2(\lambda)=0$ is a boundary of $\cT$. While the constraint on $m$ is in this case algebraic, the constraint on the original parameters $m_0$ and $\lambda$ need not be.  There is no {\it a priori}
reason for the function $m^2(\lambda)$ to be algebraic and in the 0d $\phi^4$ example it is explicitly non-algebraic.  In this case the theory space (for instability) or the correlation functions are not be $\bbR_{\rm alg}$-definable. 

While the definitions make sense for any set of $d$-dimensional QFTs, we can consider sets of QFTs with 
specific properties such as
global symmetry, conformal symmetry and/or supersymmetry.
These choices will be referred to as the ``type'' of the QFT. In order to have 
a simplified notation, we sometimes write 
\beq
   \bbR_{\text{QFT}d}\ , \qquad  \bbR_{\text{CFT}d}\ , \qquad \bbR_{\text{AQFT}d} \ , 
\eeq
in case we want to only highlight certain aspects of the structure. Here 
AQFT$d$ indicates that we require the considered QFTs to be asymptotically free. Of course, this simplified notation will not capture all the needed information and is more a placeholder for all the structures associated to such theories.  
In practice, especially when evaluating tameness, it is crucial to specify a proper 
definition of $\cT$ and $\cS$. 
The important point is that we want to make $\bbR_{\cT,\cS}$ rich enough to formulate physical questions about the theories $\cT$, but small enough such that $\bbR_{\cT,\cS} $ could be o-minimal. 
\subsubsection*{An example: quantum mechanics }

A simple example in which $\bbR^{\rm def}_{\cT,\cS}$ and $\bbR_{\cT,\cS}$ can be identified are quantum mechanical
systems with finitely many states.
So far we have mostly used the parameters $\lambda$ of the Lagrangian to parameterize the theory space. In this case we find it more convenient to define the theory by choosing a basis $\ket{i}$ for the states and specifying
the matrix elements for some finite list of operators $\cO_a$, $a_{i,j}=\langle i | \cO_a | j \rangle$, as well as the energies $\Delta_i$ of the states. Taking a single operator $\cO$ (the generalization is obvious),
the theories are parameterized by the set $\cT=\{a_{i,j},\Delta_i\}$, 
a finite dimensional parameter space. 

Using these we define time ordered correlation functions as sums over intermediate states,
\begin{equation}
      \langle \cO_1(t_1) %\cO_2(t_2) 
      \ldots \cO_n(t_n) \rangle = 
    \sum_{i} \langle 0 | \cO_1(t_1) | i \rangle e^{-{\Delta_i|t_1-t_2|}} 
    \langle i |\cO_2(t_2) \ldots \cO_n(t_n) | 0\rangle  \;,
    \end{equation}
where we consider $t_1>t_2>\ldots>t_n$.
After the insertion of sufficiently many intermediate states this reduces to a finite sum whose terms are polynomial in the matrix elements and exponential in the energies $\Delta_i$.  There is an evident generalization to the partition function using the trace.
Thus we can regard the space of quantum mechanical theories with finite dimensional Hilbert space
as parameterized by $\Delta_i$ and the matrix elements.  As these are simply real variables, if we
start with $\bbR^{\rm def}_{\cT,\cS}=\bbR_{\rm alg}$, we can define every correlation function in terms of finite sums and
products of these parameters and the exponential function.  Thus the resulting structure is $\bbR_{\cT,\cS}=\bbR_{\rm exp}$.

For quantum mechanics with an infinite dimensional Hilbert space the situation is more complicated, as the sum over the intermediate states is an infinite summation.  Nevertheless correlation functions can still be tame.
We give two simple examples, the free particle and the harmonic oscillator. Consider the transition function $\bra{x}e^{-H t}\ket{y}$. For the free particle of mass $m$ this is
\begin{equation}
\bra{x}e^{-(t_2-t_1) H}\ket{y}= \sqrt{\frac{m}{2 \pi  (t_2-t_1)}}e^{-\frac{ m (x-y)^2}{2(t_2-t_1)}}\;.
\end{equation}
For the harmonic oscillator with frequency $\omega$, it is the Mehler kernel
\begin{equation}
   \bra{x}e^{-(t_2-t_1) H}\ket{y}= \sqrt{\frac{m\omega}{2 \pi  \sinh (\omega (t_2-t_1))}}e^{-\frac{ m \omega}{2 \text{sinh}( \omega (t_2-t_1))}  \left(\cosh (\omega (t_2-t_1)) \left(x^2+y^2\right)-2 x
   y\right)}\;.
\end{equation}
In both cases the dependence on the parameters $m$ and $\omega$ and coordinate difference $t_2-t_1$ is definable in $\mathbb{R}_{\rm exp}$. Similarly, if one works in the Fock space, the 2-point correlation function of two Heisenberg ground states is given by 
\begin{equation}
    \bra{0}_{t_2}\ket{0}_{t_1}=\frac{1}{2 m \omega}e^{\omega (t_2-t_1)}\;.
\end{equation}
Inserting more operators does not add any new functional dependence, as by Wicks theorem any $n$-point function can be written as sums of products of two-point functions. E.g. for the 4-point function
\begin{align}
 \bra{0}_{t_2} x(t_4)x(t_3)\ket{0}_{t_1}=&\sum_n\bra{0}_{t_2} x(t_4)\ket{n}_{t_0}\bra{n}_{t_0}x(t_3)\ket{0}_{t_1}\\=&\frac{1}{2 m \omega}e^{-\frac{1}{2}\omega (t_2-t_1)}e^{-\omega (t_4-t_3)}
\end{align}
Thus the structure generated by Euclidean quantum mechanics of a free particle or a harmonic oscillator is still $\mathbb{R}_{\rm exp}$.

\subsection{Subtleties in the definitions}

While our definitions can be readily applied to many examples, there are several subtleties that can occur in general physical configurations. In the following we want to highlight a number of points which should be treated more carefully to make this definition
fully precise and which we postpone for part III.

\subsubsection*{Definability of the spacetime $\Sigma$} \label{sec:definable-spacetime}

In the construction of the structure $\bbR_{\cT,\cS}$ we required that the 
spacetime $\Sigma$ and its metric $g$ are definable in $\bbR^{\rm def}_{\cT,\cS}$. For 
$\bbR^{\rm def}_{\cT,\cS}$ and $\bbR_{\cT,\cS}$ to be o-minimal, every $\Sigma$ should be a tame set with metric $g$ a tame function. 
We already noted that we require $\Sigma$ to be Riemannian, but which manifolds $\Sigma$ should we allow? 

To answer this question in generality, we would need to develop a theory of tame differential geometry. 
Formally, it is not hard to state the initial definitions:
a tame manifold, as mentioned in \S \ref{sec:structurelogic}, has a definable atlas (covering $\Sigma$ with finitely many charts) and definable transition functions. A  Riemannian metric is a section of the square of the tangent bundle of $\Sigma$. 
We require that this bundle is tame, i.e.~admits finitely many trivializations and definable transition functions. The metric coefficients should then transform definably when changing charts. While it is easy to state these definitions, we are not aware of strong theorems that are implied by tameness. However, it is expected that tameness is a non-trivial constraint. Manifolds definable in $\bbR_{\rm alg}$ are the natural objects of real algebraic geometry and every o-minimal structure gives a generalization.  

Our pragmatic way to bypass 
tame differential geometry will be to 
propose some concrete choices for $\Sigma$ and its Riemannian metric $g$.
As an initial proposal we allow manifolds of topology $\bbR^k$, $S^k$, $T^k$
and direct products of these.  For a given $d$ this is a finite list of topologies.
It includes $\bbR^d$, so that we can make contact with our discussion of perturbation theory in part I, and it also includes the cases of most interest for CFTs.
For $g$,
we allow general conformally flat metrics and parameterized families of such metrics.
For example, $T^d$ has a family of flat metrics parameterized by positive definite
symmetric $d\times d$ matrices (more precisely equivalence classes under $SL(d,\bbZ)$) and $\cS=SL(d,\bbZ)\backslash SL(d, \bbR)/ SO(d)$ is coset space definable in $\bbR_{\rm alg}$ \cite{bakker2020tame}.
More general conformally flat metrics are included to allow relating cases such as
$\bbR^d$ and $S^d$, and also with an eye towards AdS/CFT dualities.

Let us stress that even with the restriction that $(\Sigma,g)$ is conformally flat, 
the definability of $(\Sigma,g)$ in an o-minimal $\bbR_{\cT,\cS}^{\rm def}$ gives additional conditions. Clearly, 
geometries $\bbR^d$, $S^d$, and $T^d$ with their standard metric are $\bbR_{\rm alg}$-definable. However, if we consider a conformal factor $\sigma$ that is not definable in $\bbR_{\rm alg}$ we potentially need to introduce a new function to $\bbR_{\cT,\cS}^{\rm def}$. This factor $\sigma$ will also 
enter the correlation functions, which shows that self-consistency of our construction 
requires it to be in 
$\bbR_{\cT,\cS}$.\footnote{For example, in $d=2$ CFTs the dependence of a correlation function on $\sigma$ is universal \cite{polyakov1981quantum}; 
it is given by the Polyakov action multiplied by a factor for each operator,
$Z(\sigma) = Z(0) e^{cS_{\rm Polyakov}[\sigma]} \prod_i e^{\Delta_i\sigma(x_i)}$.
This particular factor is removed by the canonical normalization, however.}
Hence, in view of our goal to identify o-minimal structures $\bbR_{\cT,\cS}$ constructed from correlation functions, we need to demand that $\sigma$ is $\bbR_{\cT,\cS}$-definable and adds no new non-tame function to $\bbR_{\cT,\cS}$.\footnote{We want the volume of $(\Sigma,g)$ (as a function on $\cS$), distances between pairs
of points, and other relevant geometric quantities such as cross ratios, to be definable
in $R_{\cT,\cS}^{\rm def}$. We believe this is the case for the choices above.}

We chose the dependence on operator locations to be in position space (points in $\Sigma$)
as this makes sense for any manifold $\Sigma$.  If $\Sigma$ has translation symmetry
then one could replace this with the dependence on momentum space (by taking the Fourier transform of the correlation function), and this is the definition we used in part I, \S 3.  While the two spaces of functions are related, it is not true in general that they define the same structure. In fact, it is not hard to see that a Fourier transform of a tame function, say of $\bbR_{\rm an,exp}$, can require the extension of the structure to include periodic functions (see \cite{cluckers2018integration} for recent general results). In this case the resulting structure has no chance to be o-minimal. Of course QFT correlation functions satisfy many constraints beyond those
of $\bbR_{\rm an,exp}$ and it could be that in this case the momentum and position spaces do define 
the same structure, at least for certain choices of $\Sigma$. We leave this question for future work.
In any case the position space definition is the relevant one here.

\subsubsection*{Global considerations }

In our definition of $\bbR_{\cT,\cS}$ we largely resort to a local perspective both for the theory space $\cT$ and the spacetimes in $\cS$. However, there are several interesting global aspects that need to be addressed to complete our construction in general settings. 

First, even if we restrict $\cS$ to be conformally flat metrics on $\bbR^k$, $S^k$, $T^k$
and direct products of these, to be completely precise we should say what sort of conformal boundaries we allow.
Another consideration influencing the choice of $\cS$ would be to ensure that the functions
obtainable as Green's functions are definable functions of position in a simple o-minimal structure.
This might favor restricting $\cS$ still further.

Second, correlation functions of non-scalar operators are not functions on $\Sigma$.
To keep the discussion simple, we could make them real valued by postulating frames. 
Alternatively, one could investigate the transition functions 
of the bundles and construct $\bbR_{\cT,\cS}$ such that they become definable. 

Third, we need the correlation functions of local operators as functions of $\cT$, the parameters of the 
set of QFTs.  As is familiar from discussions of renormalization, it is not entirely trivial to define a
local operator as a function of couplings.  Even in quantum mechanics, there can be ambiguities arising
from level crossings and monodromies.  One might find that on traversing a loop in $\cT$, an operator $O_1$
turns into another operator $O_2$.  While these make it difficult to define continuous functions on $\cT$, to
define a structure we do not need to do this.  It would suffice to decompose $\cT$ into finitely many cells
(using definable boundaries) such that the correlation functions are well defined in each cell.

The final issue arises if we want to discuss the set of all QFTs of a given type and is more challenging.
There is no reason to think that such a discussion could be based on a single definition of QFT.  For example while
Lagrangians provide very concrete parameter spaces, there are many QFTs with no known Lagrangian definition.
The evident way to deal with this is to define a structure for each of the definitions, call these $\bbR_{\cT_1}$,
$\bbR_{\cT_2}$ and so on.  We could then take the union of these to get a structure including all of their
correlation functions.  This is fine for the dependence of these functions on $\Sigma$, but it is possible
that definitions $\cT_1$ and $\cT_2$ will parameterize overlapping subsets of QFTs in different ways which are
not related by a definable map.  If so then $\bbR_{\cT_1}$ and $\bbR_{\cT_2}$ might lead to different structures,
and even if they are both tame the union might not be tame.  Another issue is the treatment of equivalences between
different QFTs such as dualities -- if a single QFT is represented by infinitely many points in $\cT$ then this
definition will not work (one would need to do the quotient explicitly).  We do not see these issues as leading
to essential difficulties but we postpone them for future work.

\subsection{Structures for theories with gravity} \label{sec:definable-QG}

We now briefly turn to the discussion of theories that contain dynamical gravity. 
At the classical level the most relevant theory to understand is Einstein's theory of general relativity. One might again hope to associate a structure to this theory in which all of its physical statements become definable. For example, one could try to construct this structure by adding all solutions to this classical theory to $\bbR_{\rm alg}$. However, a priori, there are no tameness conditions for solutions to Einstein's equations, since second-order differential equations do not necessarily preserve tameness. Nevertheless, solutions periodic in time can be ruled out for asymptotically flat spacetimes in pure gravity with vanishing cosmological constant \cite{Alexakis:2015ara} as well as for cosmological, i.e.~spatial compact, spacetimes \cite{galloway1984}. A periodic solution would correspond to a discrete isometry of spacetime, and these turn out to be induced by continuous isometries. This is in perfect agreement with the expectation of tameness. On the other hand, adding a negative cosmological constant allows for the construction of time-periodic solutions \cite{Chrusciel:2017yhy}.  We thus face similar challenges as in \S \ref{sec:definable-spacetime} in which we discussed the $\bbR_{\cT,\cS}^{\rm def}$-definability of the background $(\Sigma,g)$. While a comprehensive treatment of gravitational theories is beyond the scope of this work, we will collect a few  relevant observations in the following.

\subsubsection*{Fluctuating gravitation backgrounds}
It is difficult to formulate a general quantum theory of a fluctuating metric. Instead it is useful to formulate the theory perturbatively around a chosen fixed background $g$. One then studies small perturbations $\delta g$ around this background and writes down an effective theory for the field $\delta g$. In the case of Einstein gravity the resulting theory is 1-loop renormalizable. Thus if the classical background is tame, the results of part I would apply to this system and the theory would be definable in $\mathbb{R}_{\text{an,exp}}$. In the case of maximal supergravity in four dimensions this renormalizability holds at least up to seven loops.

\subsubsection*{Quantum theories of gravity and the structure $\bbR_{\rm QG}$}

Certainly the most exciting task is to investigate the space of quantum gravity theories $\cT_{\rm QG}$ and the associated structure $\bbR_{\rm QG}$. 
This requires one to face multiple challenges and we want to highlight a few in the following. 

We do know very little about quantum gravity in the needed generality to define $\cT_{\rm QG}$. Much work has gone in understanding string theory as a promising candidate. Trusting the lessons learned over the last decades one might hope that actually $\cT_{\rm QG}$ is just a single theory, M-theory. Unfortunately, we have no general formulation of this theory, apart from its incarnations in various limits.

Accepting this significant limitation, one might nevertheless try to construct part of $\bbR_{\rm QG}$. Here one faces another issue, since it is well-known that it makes not much sense to construct $\bbR_{\rm QG}$ from correlation functions of local operators due to the diffeomorphism symmetry of gravity.
One could bypass this by constructing $\bbR_{\rm QG}$ from the S-matrix instead, as it is done in the study of string scattering amplitudes. 
We might then ask if $\bbR_{\rm QG}$ defined by purely perturbative information can be expected to be o-minimal. However, one can see at the example of perturbative string theory that this is not the case. In fact, if we determine the associated structure $\bbR_{\rm pString}$ from string scattering amplitudes, it is easy to see that
the presence of infinitely many poles in these amplitudes implies that  
$\bbR_{\rm pString}$ cannot be o-minimal. 
However, as already mentioned in part~I, one can hope that this is an artefact of the perturbative treatment. Adding an additional direction to the string will likely make the spectrum continuous and the associated functions have better chances to be tame.

\subsubsection*{String/M compactification to AdS 
and the structure $\bbR_{\rm AdS}$}

The best definition we have of quantum gravity at present uses the AdS/CFT correspondence \cite{maldacena1999large}.
As discussed in reviews \cite{aharony_large_1999}, an anti-de Sitter spacetime $M_{d+1}$ has a conformal boundary $\Sigma$, and observables in quantum gravity on $M_{d+1}$ correspond to observables in a CFT$d$ on $\Sigma$.  Thus, it would seem, we have already defined the relevant structure for these theories: it is  $\bbR_{{\rm CFT}d}$.

On reflection, it is not obvious that this is the definition of quantum gravity we want.
The issue is that the set of quantum gravity theories we get this way includes all sorts of theories which look very different than what we need to describe the real world.  It includes theories with very small AdS radius (very large negative cosmological constant) and higher spin gravity theories \cite{giombi_tasi_2016}.  And this only scratches the surface -- the landscape of CFT2 and CFT3 (with no bound on the central charge) seems far larger than any known construction of string/M  theory compactifications would produce.

This motivates defining subsets of CFT$d$ which satisfy additional constraints motivated by gauge-gravity duality, especially the constraint that the gravity dual is local.\footnote{ 
More precisely, while the interactions can be spread over a length scale $l_s$ such as the string scale, this is much less than the AdS radius, $l_s \ll R_{AdS}$. }
Such a proposal was made in Ref. \cite{heemskerk_holography_2009}, which argues that any CFT with a large N expansion and for which all single trace operators of spin greater than 2 have (parametrically) large dimensions, has a local dual.  This proposal was developed in many works, see \cite{caron-huot_ads_2021} and references there.

It would be interesting to define a subset of CFT$d$'s satisfying these conditions
and a corresponding structure.  Since the conditions of \cite{heemskerk_holography_2009}
are parametric, we presumably would do this by defining subsets satisfying sharp
bounds and then showing that the resulting structures do not depend on the specific
bounds.  However, the conditions are not formulated in terms of individual QFTs
but rather families of QFTs (say depending on $N$), so they go beyond the definitions
made here.  We leave a precise definition for future work.

On the dual side, we could use the correspondence for correlation functions 
\cite{aharony_large_1999} to express the large $N$ limit of general CFT correlation
functions in terms of classical supergravity calculations.  Classical general relativity and
even supergravity are much better understood mathematically than QFT, and working with
this structure and checking its tameness  would be very interesting even if restricted to the pure gravity sector.  Tameness of the space of these theories is a refinement of the 
finiteness conjecture of \cite{acharya_finite_2006} and could be studied as well.

\section{Tameness of physical observables} \label{sec:tameobservables}

Having introduced structures $\bbR_{\cT,\cS}^{\rm def}$ and $\bbR_{\cT,\cS}$ associated to 
a set of QFTs $\cT$ and a set of spacetimes $\cS$, we now 
want to understand when they are o-minimal. To do this, (1) we need to ensure that $\bbR_{\cT,\cS}^{\rm def}$, in which $\cT$ and $\cS$ are definable, is an o-minimal structure, and (2) we need to 
ensure that adding the observables to $\bbR_{\cT,\cS}^{\rm def}$ as discussed in \S \ref{sec:structureQFT} preserves tameness. 
Let us postpone the specification of a tame set of theories $\cT$ with as few constraints as possible to \S \ref{Tameness-cT}. 
Here we will focus on the second step: given that $\bbR_{\cT,\cS}^{\rm def}$ is o-minimal,
let us discuss criteria that could ensure that $\bbR_{\cT,\cS}$ is o-minimal.  

\subsection{QFTs on points} \label{sec:QFTs_on_points}

Let us begin with the simplest case, a QFT in zero dimensions, {\it i.e.}~on a point $p$. We have 
already mentioned the $\phi^4$-theory in part I, but in the following 
we want to be more general. Therefore, let us assume that $S^{(0)}$ 
is the action of such a QFT with finitely many scalar fields $\phi_1,...,\phi_n$. 
While in higher dimensions, $S^{(0)}$ is an integral over 
spacetime, in zero dimensions this is simply a function of fields evaluated at the point $p$. The set of spacetimes $\cS=\{p\}$ is therefor tame. Let us take the dependence of the Lagrangian on $\phi$ and parameters $\lambda$ to be definable in some o-minimal structure $\bbR^{\rm def}_{\cT}$,
and let us consider correlation functions of operators $O_k$ which are also $\bbR^{\rm def}_{\cT}$-definable functions of $\phi$.
The simplest case would be polynomial operators $O_k=\phi^k$ and $S^{(0)}=\sum_k \lambda_k O_k$ but one can be more general.

In zero dimensions, the path integral \eqref{correlation_fct_x} reduces to an ordinary integral
\beq \label{0dintegral}
   \langle \cO_1 ... \cO_n  \rangle_\lambda  = \int d\phi_1... d\phi_k \, \cO_1 ... \cO_n\,   e^{-S^{(0)}(\phi,\lambda)} \ , 
\eeq
where the integration is performed over the space of field configurations. 

Now, we can ask
 (1) Is the structure $\bbR_{\cT,\text{exp}}^{\rm def}$ obtained by extending $\bbR_{\cT}^{\rm def}$ by the exponential map an o-minimal structure? 
 (2) Does the $\bbR_{\cT,\text{exp}}^{\rm def}$-definable 
function in \eqref{0dintegral} integrate to a $\lambda$-dependent function that is 
definable in some o-minimal structure $\bbR_{\rm QFT0}$?

The first question can be answered affirmatively as shown in \cite{9710220}.\footnote{Note that the 
exponential function is always definable in the Pfaffian closure of $\bbR_{\cT}^{\rm def}$.}
Every o-minimal structure remains o-minimal if one adds the real 
exponential function. A simple example of this general fact is given in \S \ref{sec:o-minimalstructures} where we listed both the o-minimal structures $\bbR_{\rm an}$ and $\bbR_{\rm an,exp}$.

The second question on whether integrals of tame functions are also tame is longstanding in mathematics and has not been answered conclusively yet. There are, however, several interesting results to mention in this context.\footnote{We are grateful to Tobias Kaiser for useful correspondence on this topic.} Firstly, it is generally understood that if a real-valued function $f(\phi,\lambda)$ is definable in $\bbR_{\rm an}$ then its 
integral $g(\lambda)=\int d^k\phi\, f(\phi,\lambda)$ is definable in $\bbR_{\rm an,exp}$ \cite{AIF1998}. This general result naturally relates to the fact that Feynman integrals are definable as discussed 
in part I.\footnote{We note that it is not immediate that it implies the tameness of Feynman integrals, since often one needs to remove singularities of the integral by regularization and renormalization. This leads one away from ordinary integrals over real functions and the connection with period integrals becomes relevant.} In fact, this result is rather generally applicable in perturbative expansions in which the exponential in \eqref{0dintegral} turns into a polynomial
expression. When including the exponential function only partial results are known. One important general result of \cite{Speissegger1997ThePC} is that if $f:\bbR \rightarrow \bbR$ 
is definable in some o-minimal structure $\widehat \bbR^{\rm def}$ then $\int_0^x f(t)dt$ is definable in the Pfaffian closure $\mathcal{P}(\widehat \bbR^{\rm def})$ of $\widehat \bbR^{\rm def}$, as introduced in \S \ref{sec:o-minimalstructures}. Certain special integrals with exponential functions have also been shown to be tame in \cite{KaiserRuppert} and special  tame measures have been proposed in \cite{KaiserMeasure}. 

The general statement about the existence of an o-minimal structure $\bbR_{\rm QFT0}$ follows from the following conjecture put forward in \cite{vandenDriesConj,KAISER20121903}.

\noindent
\textbf{\linkdest{Conjecture1}Conjecture 1.} \textit{For every o-minimal structure $\widehat \bbR^{\rm def}$ there exists an o-minimal structure $\widehat \bbR $ such that for 
every real-valued function $f(\phi,\lambda)$ that is definable in  $\widehat \bbR^{\rm def}$ and every set $D$ definable in $\widehat \bbR^{\rm def}$ the following holds: \\[.1cm]
(1) The set $\big\{\lambda \in \bbR^m: 
\int_{D} d^k\phi |f(\phi,\lambda)| < \infty \big\}$ is definable in $\widehat \bbR$.\\[.1cm]
(2) The function $g(\lambda)=\int_D d^k\phi\, f(\phi,\lambda)$ is definable in $\widehat \bbR$.}

\noindent
While this conjecture is formulated in flat space, one can extend it to definable manifolds and include definable measures.

Let us note that the above considerations naturally extend 
to higher dimensional QFTs localized on $N$ points, e.g.~forming a finite lattice. 
Such settings were studied recently in~\cite{Gasparotto:2022mmp}. Consider a QFT with a single 
field $\phi$ and localize it on $N$ spacetime points $y_i \in \Sigma$ by setting $\phi_i \equiv \phi(y_i)$.\footnote{The generalization to finitely many fields is straightforward.} Reducing the action of the higher-dimensional theory to these finitely many points will lead to an (approximate) zero-dimensional action $S^{(0)}(\phi_i,\lambda)$. 
In ref.~\cite{Gasparotto:2022mmp} the integrals 
\beq
   I_{\nu_1...\nu_N}(\lambda) = \int_{\cC^N} d^N \phi\, \prod_{i}^N \phi_i^{\nu_i} \, e^{-S^{(0)}(\phi,\lambda)}\ , 
\eeq
have been studied to determine the $\lambda$-dependence of correlation functions polynomial in the $\phi_i$. Clearly, by \hyperlink{Conjecture1}{Conjecture 1}, these integrals are definable in some o-minimal structure if $S^{(0)}(\phi,\lambda)$ is. 
In fact, it was shown in \cite{Gasparotto:2022mmp} that for simple polynomial $S^{(0)}(\phi,\lambda)$ the $I_{\nu_1...\nu_N}$ satisfy a set of differential equations that are solved by products of rational functions and generalized hypergeometric functions in $\lambda$. The appearing hypergeometric functions are of type $_pF_p$ and the challenge to establish their tameness has been pointed out already in part I. 

\subsection{Simple observables in higher-dimensional QFTs}

For general higher-dimensional QFTs we do not know of mathematical theorems or conjectures which imply the tameness of physical observables. Recall that 
we added correlations functions in moving from $\bbR^{\rm def}_{\cT,\cS}$ to $\bbR_{\cT,\cS}$ that are, as indicated in \eqref{correlation_fct_x}, given by path integrals over field configurations. The study of such integrals is very involved and has been, to our knowledge, not been undertaken in the o-minimal context even for simple examples. In part I these integrals have been studied from a perturbative approach. At this level the correlation functions are tame functions. Here we want to study the tameness of these objects at the nonperturbative level.

To gauge our expectations on the tameness of observables, let us consider the example of a Klein-Gordon field 
of mass $m$ on $d$-dimensional Euclidean space $\Sigma$.
As we are dealing with a free theory, the $n$-point correlation functions can be computed using Wick's theorem as a 
sums of products of the two-point correlation functions
\begin{equation}
    O^{(d)}(y_1,y_2)=\bra{ \mathbf{y}_2}e^{- H (t_2-t_1)}\ket{\mathbf{y}_1}\,, \qquad y_i = (t_i,\mathbf{y}_i) \in \Sigma\ . 
\end{equation}
For a Klein-Gordon field in flat space this two-point function can be evaluated as (see e.g.~\cite{Bertola:1999vq})
\begin{equation}
     O^{(d)}(y_1,y_2)=(2\pi)^{-d/2}\left(\frac{(y_2-y_1)^2}{\sqrt{m}}\right)^{\frac{d-2}{2}}K_{\frac{d-2}{2}}(\sqrt{m}(y_2-y_1)^2)\;.
\end{equation}
$K_a(x)$ denotes the modified Bessel function of the second kind. As already discussed in \S \ref{sec:QFTs_on_points}, these functions are in the generic case exponential periods and \hyperlink{Conjecture1}{Conjecture 1} implies that they are definable in some o-minimal structure.\footnote{Only in the special cases of $d=1$ or $d=3$ the function reduces to $K_{1/2}(x)=e^{-x}\sqrt{\pi/(2x)}$ and definability in $\bbR_{\rm exp}$ is apparent. It turns out that $K_\alpha(x)$ is definable in the Pfaffian structure $\mathcal{P}(\mathbb{R}_{\rm alg})$, see v2 of \cite{Douglas:2022ynw} and \cite{complexappear}.} If so, then all correlation functions of a finite number of operators are tame. 

While tameness is realized in flat space, assuming \hyperlink{Conjecture1}{Conjecture 1}, already in AdS spacetime this no longer holds. AdS space is definable in $\bbR_{\rm alg}$, but the tameness of the two-point function can no longer be established without imposing further constraints on the parameters. It is given by \cite{Fronsdal:1974ew} 
\begin{equation} \label{AdS-two-point-function}
    O^{(d)}_{{\rm AdS}_{d+1}}(y_1,y_2)=\frac{e^{-i\pi (d-1)/2}}{(2\pi)^{(d+1)/2}}((y_1\cdot y_2)^2-1)^{-\frac{d-1}{4}}Q^{(d-1)/2}_{\sqrt{d^2/4+m^2}-1/2}(y_1\cdot y_2)\ ,
\end{equation}
where $Q_a^b(x)$ denotes the Legendre function of the second kind. The only dependence on the mass is in the lower argument of the Legendre function, $\lambda=\sqrt{d^2/4+m^2}-1/2$. We can now see that the dependence on the parameter $m$ is not tame when considering all $m \in \bbR$, since the $Q_{a}^b(x)$ oscillates in $a$.

Already this simple example shows that we have to be careful when making claims about the tameness of correlation functions. We will suggest two ways how we expect that such violations can be avoided: 
\begin{itemize}
    \item[(1)] consider the case in which the theory becomes conformal, i.e.~send $\lambda \rightarrow 0$;
    \item[(2)] introduce a cut-off scale $\Lambda$ and restrict the mass $|m|<\Lambda$.
\end{itemize}
In case (1) one checks that the Legendre function reduce to an algebraic function and hence is definable in $\bbR_{\rm alg}$. In case (2) we use the analyticity of $Q$ in $a$ to infer that it is restricted analytic on $|m|<\Lambda$.\footnote{$Q_{a}^b(x)$ can be expressed in terms of Gauss hypergeometric functions, which are entire functions of its parameters if one excludes the singular points $0$, $1$ and $\infty$ for the argument.} 

While it is by no means immediate that these observations generalize easily to interacting theories, we will use them as inspiration to put forward two conjectures about the tameness of correlation functions  in effective theories compatible with quantum gravity and in CFTs in 
the following.

\subsection{Tame observables in effective field theories}

Let us first discuss the $d$-dimensional effective field theories with finitely many dynamical fields and a cut-off energy scale $\Lambda$. We will denote the space of all such theories by $\cT_{\text{EFT}d}$. Clearly, the definition of what we exactly mean by $\cT_{\text{EFT}d}$ requires much effort. A simplified way to think of this space, is to consider only Lagrangian theories and collect all Lagrangians with their field dependence and parameter dependencies. The graphs of these Lagrangians evaluated over the field space and parameter space are sets and their formal union can be thought of as being the space $\cT_{\text{EFT}d}$.  

Determining the conditions on the effective theories such that $\cT_{\text{EFT}d}$ is definable in some o-minimal structure, is  
a very non-trivial task. However, it was proposed in \cite{Grimm:2021vpn} that a sufficient condition for $\cT_{\text{EFT}d}$ to be tame is that this space contains all effective theories that can be consistently coupled to quantum gravity. This let to the conjecture: 

\noindent
\textbf{\linkdest{Conjecture2}Conjecture 2 (Tameness of EFTs consistent with quantum gravity).}
\textit{The set $\cT_{{\rm EFT}d}$ of all effective field theories  that are valid at least below a fixed cut-off scale $\Lambda$ and can be consistently coupled to quantum gravity is definable in some o-minimal structure.}

In other words, determining the necessary set of  constraints on tame $\cT_{\text{EFT}d}$ should  be viewed as being part of the swampland program, see in e.g.~\cite{Palti:2019pca,vanBeest:2021lhn} for a  survey.  Let us stress  that from a low-energy point of view, without reference to an UV completion with gravity, it is not apparent what a minimal set of conditions on can ensure that $\cT_{\text{EFT}d}$ is tame. 

Given a well-defined set of effective theories $\cT_{\text{EFT}d}$, one can again 
 construct an associated structure $\bbR_{\text{EFT}d}$, depending on some set $\cS$ of Euclidean spacetimes. To evaluate the correlation functions one now has to carefully implement the cutoff $\Lambda$, e.g.~by considering a momentum expansion and retaining terms up to a finite order in $\Lambda$. At least for the perturbative expansion we expect that the results of 
 part I can be generalized and indicate that $\bbR_{\text{EFT}d}$ is o-minimal. We believe that this property persists and formulate:
 
 \noindent
\textbf{\linkdest{Conjecture3}Conjecture 3 (Tameness of EFT correlation functions).}
\textit{The structure $\bbR_{{\rm EFT}\textit{d}}$ built from all correlation functions for the theories in $\cT_{{\rm EFT}d}$ with cut-off $\Lambda$ on tame spacetime manifolds $(\Sigma,g)$ is o-minimal.}

In general, it is very challenging to determine $\bbR_{\text{EFT}d}$ for all theories in $\cT_{\text{EFT}d}$
determined only by consistency with quantum gravity and validity below $\Lambda$. However, to test \hyperlink{Conjecture3}{Conjecture 3}  one can consider any set of effective theories determined from string theory and investigate its correlation functions on some chosen tame spacetime manifold.

\subsection{Tame observables in conformal field theories} \label{sec:tameCFTobservables}

In the following we turn our attention to CFTs in $d$ spacetime dimensions.
In contrast to the effective field theories discussed in the previous section, CFTs have no scale and therefore do not require a UV completion. In the following we will claim that the tameness of observables should therefore be a general feature of tame sets of CFTs. We formulate this expectation as 
a conjecture:

\noindent
\textbf{\linkdest{Conjecture4}Conjecture 4 - Tame CFT Observables Conjecture.} \textit{Given a tame set $\cT$ of CFTs  and a tame set $\cS$ of Riemannian manifolds, the canonically normalized partition and correlation functions computed among finite sets of tame operators are tame functions varying over $\cT$, $\cS$. 
In other words, assuming that $\cT$, $\cS$, and considered operators $\cO_i$ are definable in some o-minimal structure $\bbR^{\rm def}_{\cT,\cS}$, the above constructed structure $\bbR_{\cT,\cS}$ of physical observables is o-minimal.}

While \hyperlink{Conjecture4}{Conjecture 4} is clearly speculative, we know a lot about CFTs and their correlation functions to give non-trivial evidence. 
We will first discuss free CFTs, in which case the conjecture is most straightforwardly supported. The true power of the statement turns out to emerge for interacting CFTs and we will give a concise discussion of conformal blocks to gather some support. Remarkably, there are a number of non-trivial implications of \hyperlink{Conjecture4}{Conjecture 4}, which we will outline at the end of this section.

\subsubsection*{Free CFTs}

To begin with, let us comment on correlation functions of $n$ operators defined in a tame set $\cT$ of free CFTs. 
Using Wick's theorem they can be expressed in terms of the 2-point functions. The latter are fixed by conformal symmetry and take the form
\begin{equation} \label{two-point}
    \langle\mathcal{O}_i(x_1)\mathcal{O}_j(x_2)\rangle=\frac{\delta_{ij}}{(x_1-x_2)^{\Delta_i+\Delta_j}}\;,
\end{equation}
where $\Delta_i$ are the scaling dimensions of $\cO_i$.
These functions are both tame in the spacetime positions $x_i$ as well as in the $\Delta_i$. For fixed rational $\Delta_i$ the function \eqref{two-point} is simply algebraic in $x_i$. To also allow for variations in the exponent $\Delta_i$ one has to consider a larger o-minimal structure such as $\bbR_{\rm an,exp}$. Returning to the $n$-point correlation functions, we first note that the number of possible Wick contractions is finite. It is thus clear that each of these functions is obtained as finite sum of products of tame functions and therefore is also tame. Tameness hereby holds both in the spacetime positions as well as in the operator dimensions. Note that this is true as long as we can use a finite set of $\Delta$ as parameters of the set $\cT$ of free theories. Thus \hyperlink{Conjecture4}{Conjecture 4} holds in all examples of this type.

\subsubsection*{Interacting CFTs}

Now we turn to the case of considering a tame set $\cT$ of interacting CFTs. As in the non-interacting case, conformal symmetry strongly constrains the possible form of the correlation functions. In fact, it suffices to completely determining the spacetime dependence of the 2- and 3-point functions.
The 3-point functions are given by
\begin{equation}
    \langle\mathcal{O}_1(x_1)\mathcal{O}_2(x_2)\mathcal{O}_3(x_3)\rangle=\frac{C_{1,2,3}}{x_{12}^{\Delta_1+\Delta_2-\Delta_3}x_{23}^{-\Delta_1+\Delta_2+\Delta_3}x_{13}^{\Delta_1-\Delta_2+\Delta_3}}\;,
\end{equation}
where we abbreviated $x_{ij}=x_i-x_j$. The $C_{1,2,3}$ are the theory dependent constants that can depend on the parameters but not on the spacetime coordinates. In fact, it is natural to view $C_{1,2,3}$ as parameterizing $\cT$ directly (see e.g.~\cite{Douglas:2010ic}). The tameness in these parameters is then trivial. The tameness in the spacetime coordinates and operator dimensions follows as in \eqref{two-point}. The first non-trivial correlator is thus the 4-point function, which we discuss in the following.  

It turns out to be useful to separate the theory dependent data in terms of the 3-point coefficients and the universal structure of the conformal symmetry by introducing conformal blocks. We will first discuss the general properties of these and then specialize to the case of two-dimensional CFTs where one can understand the tameness properties. Finally we will look at some examples of simple CFTs and the structures generated by them.
The four-point correlators will depend on the positions of the four operator insertions $x_i$ only through the conformally invariant crossing ratio
\begin{equation}
 z=\frac{x_{12}x_{34}}{x_{13}x_{24}}\;.
\end{equation}
The 4-point correlator can then be decomposed into a sum over Virasoro conformal blocks $\mathcal{F}(c,\Delta_i,\Delta,z)$, which sum the contributions of the infinite descendent states.\footnote{We would like to thank Volker Schomerus 
for pointing out a mistake in the first version of this paper in which we have oversimplified the statements on conformal blocks.} Explicitly, this decomposition takes the form 
\begin{align}
\label{4point}
 \langle \mathcal{O}_1(x_1)\mathcal{O}_2(x_2)&\mathcal{O}_3(x_3)\mathcal{O}_4(x_4)\rangle=\sum_{\mathcal{O}\in \mathcal{O}_1\times\mathcal{O}_2} C_{1,2,\mathcal{O}}C_{3,4,\mathcal{O}}\mathcal{F}(c,\Delta_i,\Delta,z)\;,%\\
%    W_\mathcal{O}&=\frac{1}{C_{1,2,\mathcal{O}}C_{3,4,\mathcal{O}}}\sum_{{\alpha\in}{\rm descendants}} \bra{0}\mathcal{O}_1\mathcal{O}_2\ket{\alpha}\bra{\alpha}\mathcal{O}_3\mathcal{O}_4\ket{0}\;,
\end{align}
where $z$ is the crossing ratio and the $C_{1,2,3}$ denote the OPE coefficients which together with the operator dimensions $\Delta_i$ capture the theory-dependent information. To analyze the tameness properties of the 4-point function, we now restrict to CFTs with only finitely many primary operators, as it is the case for rational conformal field theories. In this case only finitely many conformal blocks contribute to a given 4-point correlator, from which it follow that the tameness of the conformal blocks implies the tameness of the 4-point function. If there are infinitely many primaries, the sum will be replaced by an integral and \hyperlink{Conjecture1}{Conjecture 1} of \S \ref{sec:QFTs_on_points} ensures the tameness in this case.

In general, no closed form is known for the Virasoro conformal blocks. We will focus on some examples where such a form is known, like the large $c$ limit or the minimal models. In these cases the conformal block is given by a Gauss hypergeometric $_2F_1$ function \cite{Belavin:1984vu}. To be precise, in the large $c$ limit the conformal block is given by \cite{Perlmutter:2015iya}
\begin{equation}
\label{2dpartialwave}
    \lim_{c\rightarrow \infty}\mathcal{F}(c,\Delta_i,\Delta,z)=z^\Delta\; _2F_1 \Big(\Delta+\Delta_{12},\Delta+\Delta_{34};2\Delta;z\Big)\;.
\end{equation}
This function oscillates in the differences $\Delta_{ij}=\Delta_i-\Delta_j$ of the conformal dimensions, while it behaves in a tame way in $\Delta$ and $z$. To make this behaviour explicit, we specialize to the case of unit crossing ratio:
\begin{equation}
 \mathcal{F}(\infty,\Delta_i,\Delta,1)=\frac{\pi  \csc \left(\pi  \left(-\Delta _{12}-\Delta _{34}\right)\right) \Gamma (2 \Delta )}{\Gamma \left(\Delta -\Delta _{12}\right) \Gamma \left(\Delta -\Delta _{34}\right) \Gamma \left(\Delta _{12}+\Delta _{34}+1\right)}\;.
\end{equation}
For any fixed $\Delta$ this function has infinitely many zeros located at $\Delta-\Delta_{ij}=-n$ for $n\in \mathbb{N}$ due to the poles of the gamma functions in the denominator.
%on the right hand side is proportional to $\text{Sinc}^2(\pi \Delta_{1,2})$. As sinc is a non-tame function for unbounded argument
This appears to be in conflict with the tameness of the correlation function as demanded by \hyperlink{Conjecture4}{Conjecture 4}. Note that the dependence on $\Delta$ is unproblematic, as in an unitary CFT the dimensions are non-negative and the $\Gamma$-function on the positive real line can be defined in the o-minimal structure $\mathbb{R}_{\mathcal{G},\text{exp}}$ as we recalled  in \S \ref{sec:o-minimalstructures}. We will return to this issue in \ref{sec45}. There are three possibilities to restore the tameness of the 4-point functions. First, the coefficients of the conformal blocks, i.e. the fusion coefficients, could be vanishing. This is for example the case in free theories. Second, there could be cancellations between the different blocks. This is especially a possibility if there are infinitely many states and thus infinitely many contributions to the amplitude. Finally, the differences of the operator dimensions could be bounded. Turning this argument around, demanding tameness of the correlation functions as in \hyperlink{Conjecture4}{Conjecture 4} requires that the differences of the operator dimensions is bounded in interacting CFTs with finitely many primary states. 

We like to stress that this is just the leading behaviour of the conformal blocks in the $c\rightarrow\infty$ limit. In the general case a closed form expression is unknown. Usual approaches to compute the conformal blocks are either recursive or in terms of so-called Dotsenko-Fateev integrals introduced in the seminal paper \cite{Dotsenko:1984nm}. These integrals are generalizations of Euler type integrals. For the 4-point function on the sphere they take the form \cite{Mironov:2010zs}
\begin{equation*}
    Z=z^{\frac{\alpha_1\alpha_2}{2\beta}}(1-z)^{\frac{\alpha_2\alpha_3}{2\beta}}\prod_{i=1}^{N_1}\int_{0}^z\!\!\! \mathrm{d}x_i \!\!\prod_{i=N_1+1}^{N_1+N_2}\int_0^1\!\! \! \mathrm{d}x_i\prod_{i<j}(x_i-x_j)^{2\beta}\prod_i x_i^{\alpha_1}(x_i-z)^{\alpha_2}(x_i-1)^{\alpha_3}.
\end{equation*}
The parameters $\alpha_i$, $\beta$ and $N_i$ are rational functions of the parameters of the theory, for the exact dependence we refer to \cite{Mironov:2010zs}. This integral is a period and thus tame in the crossing ratio $z$, while the tameness in the remaining parameters is not clear. It would be interesting to study this in more detail.

\subsubsection*{An example CFT with a parameter: Liouville theory}

An example of an interacting theory with a free parameter is the Liouville theory. The Liouville theory is defined on a two dimensional surface with metric $g_{ab}$ and curvature $R$ by the Lagrangian
\begin{equation}
    \mathcal{L}=\frac{1}{4}g^{ab}\partial_a\phi \partial_b \phi +\mu e^{2b \phi}+\frac{Q}{4\pi}R \phi\;.
\end{equation}
On a fixed geometry this Lagrange depends on two parameters, the background charge $Q$ and the coupling constant $b$. The  parameter $\mu$ can be freely chosen to take any non-zero value and has no physical effect as it can be absorbed by a constant field redefinition. The theory is a CFT if these parameters fulfill the relation
\begin{equation}
    Q=b+\frac{1}{b}\;.
\end{equation}
In this case the theory is an example of an interacting CFT which has one free parameter. As the central charge $c= 1+6Q^2$ is fixed by the choice of the background charge $Q$ the upper bound on $c$ restricts the allowed values of $Q$ to a compact interval. The primary fields are given by the exponentials $\mathcal{O}_{\alpha}=e^{\alpha\phi}$ of conformal dimension $\Delta_{\alpha}=\alpha(Q-\alpha)$. The relevant operators have 
\begin{equation} \label{def-alpha-Delta}
    \alpha=Q/2+i p\ , \qquad \Delta=p^2+Q^2/4\ .
\end{equation}
Due to the simplicity of the theory the 3-point structure constants as well as the 4-point correlators can be computed explicitly.  The structure constants are given by the DOZZ formula \cite{Dorn:1994xn,Zamolodchikov:1995aa}
\begin{equation}
    C_{\alpha_1,\alpha_2,\alpha_3}=\left(\pi \mu \gamma(b^2)b^{2-2b^2}\right)^{\frac{Q-\alpha}{b}}\frac{\Upsilon(b)\Upsilon(2\alpha_1)\Upsilon(2\alpha_2)\Upsilon(2\alpha_3)}{\Upsilon(\alpha-Q)\Upsilon(\alpha-2\alpha_1)\Upsilon(\alpha-2\alpha_2)\Upsilon(\alpha-2\alpha_3)}\;,
\end{equation}
where $\alpha=\alpha_1+\alpha_2+\alpha_3$ and $\gamma(x)=\Gamma(x)/\Gamma(1-x)$. The Upsilon function $\Upsilon(x)$ is an entire function with the integral representation
\begin{equation}
    \Upsilon(x)=\exp\left[\int_{0}^{\infty}\frac{{\rm{d}}t}{t}\left(\big(\tfrac12 Q -x\big)^2e^{-t}-\frac{\sinh^2(\frac12 Q -x)\frac{t}{2}}{\sinh(bt/2)\sinh(t/2b)}\right)\right]\;.
\end{equation}
This function is a cousin of the usual $\Gamma$ function. Note that the only zeroes of the function are the infinitely many zeroes along the real axis and it has no poles. The $\Upsilon$ function is tame in the imaginary direction and an analytic function in the real direction. In the DOZZ formula the arguments of the $\Upsilon$ function are of the form $Q/2+ip$. If we consider theories $\cT$ of central charge $c=1+6Q^2$ bounded by some fixed constant, as we will suggest is necessary for the tameness of $\cT$ in \S \ref{Tameness-cT}, $Q$ is restricted to an interval. This ensures that the structure constants are tame both in $Q$ and $p$.  

The 4-point correlator on the sphere can now be expressed in terms of the coefficients $C_{\alpha_1,\alpha_2,\alpha_3}$ and the hypergeometric conformal blocks as 
\begin{equation}
    \langle \mathcal{O}_{\frac{-1}{2b}} \mathcal{O}_{\alpha_1}\mathcal{O}_{\alpha_2}\mathcal{O}_{\alpha_3}\rangle= \int_{0}^{\infty}{\rm{d}}p \, C_{-\frac{1}{2b},\alpha_1,\frac{Q}{2}+ip}C_{\frac{Q}{2}+ip,\alpha_3,\alpha_4}\mathcal{F}(c,\Delta_i,\Delta,z)\ ,
\end{equation}
where $\Delta$ is given in \eqref{def-alpha-Delta}.
As there are infinitely many primaries in this case the sum of \eqref{4point} is replaced by an integral over the momenta $p$ labeling all the primaries. All functions appearing in this integral are tame functions of $p$, thus \hyperlink{Conjecture1}{Conjecture 1} of \S \ref{sec:QFTs_on_points} implies that the 4-point correlators are also tame functions.

Let us close by noting that an alternative route to collect evidence for \hyperlink{Conjecture4}{Conjecture 4} is to exploit the AdS/CFT duality. Here multiple new challenges arise, since the identification of parameters on the two sides of the duality can be non-trivial. For example, consider the Klein-Gordon field in AdS$_d$
with two-point function \eqref{AdS-two-point-function}. By duality the mass of the particle corresponds to the operator dimension. Furthermore, this operator dimension is a natural parameter of the CFT. It therefore appears that the non-tameness of  \eqref{AdS-two-point-function} in the mass gives 
a counterexample to \hyperlink{Conjecture4}{Conjecture 4}. However, a closer inspection shows that one cannot simply take the limit $m\rightarrow\infty$ for a particle in AdS space, since we expect the theory to break down when the Schwarzschild radius of the particle exceeds the radius of the AdS space. On the AdS side this corresponds to a natural cutoff of the theory, but we could scale both the mass $m$ and the AdS radius. 
It turns out that on the CFT side this leads to a bounded operator dimension and hence tameness is not violated. This is clearly just the beginning of a potentially interesting story that should be explored in the future.

\subsection{Previous conjectures related to Conjecture 4}
\label{sec45}
Let us now suppose that \hyperlink{Conjecture4}{Conjecture 4} is true and denote that o-minimal structure  by $\bbR_{\text{CFT}d}$. 
While it is beyond the scope of this work to systematically study implications, 
let us make some remarks, on separation of conformal dimensions, and on
the convergence of conformal perturbation theory. 

\subsubsection*{No parametric separation of operator dimensions}

A first danger to the validity of \hyperlink{Conjecture4}{Conjecture 4} has already appeared in \S \ref{sec:tameCFTobservables}.
We saw that it can be violated if it is possible within a set of CFTs to 
send differences $\Delta_{12}= \Delta_1 - \Delta_2$ between conformal dimensions of interacting operators\footnote{Interacting here means the existence of a non-vanishing fusion coefficient $C_{\cO_1,\cO_2,\cO_3}$ with $\cO_3\neq 1$.} $\cO_1,\cO_2$ parametrically to infinity.  Suppose this were the case, then individual conformal blocks
involving these operators would be
non-tame functions of the parameter.\footnote{In two dimensions and large central charge, for example, the relevant function is
$_2F_1(x,y,z;u)$, appearing in \eqref{2dpartialwave}, which is oscillating in $x$ for fixed $y,z,u$.  }
If the full correlation function (summing over all conformal blocks) is also non-tame,
this would contradict \hyperlink{Conjecture4}{Conjecture 4}.  One can imagine ways out of this,
for example a sum over finely or infinitely many non-tame conformal blocks could produce a tame function,
but it is suggestive. In particular, in case one has only 
finitely many conformal blocks, there must be an underlying reason such that the non-tame contributions cancel. 

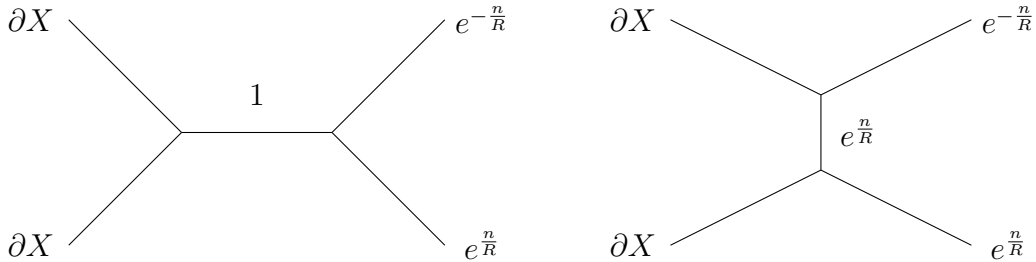
\begin{figure}[h]
    \centering
    \begin{tikzpicture}
\draw (-2.5,0)--(-1,1.5);
\draw (-2.5,3)--(-1,1.5);
\draw (-1,1.5)--(1,1.5);
\draw (1,1.5)--(2.5,3);
\draw (1,1.5)--(2.5,0);
\node at (-3,0) {$\partial X$};
\node at (-3,3) {$\partial X$};
\node at (3,0) {$e^{\frac{n}{R}}$};
\node at (3,3) {$e^{-\frac{n}{R}}$};
\node at (0,2) {$1$};

\draw (5.5,0)--(7.5,1);
\draw (5.5,3)--(7.5,2);
\draw (7.5,2)--(7.5,1);
\draw (7.5,2)--(9.5,3);
\draw (7.5,1)--(9.5,0);
\node at (5,0) {$\partial X$};
\node at (5,3) {$\partial X$};
\node at (10,0) {$e^{\frac{n}{R}}$};
\node at (10,3) {$e^{-\frac{n}{R}}$};
\node at (8,1.5) {$e^{\frac{n}{R}}$};
\end{tikzpicture}
    \caption{The two different ways to expand the 4-point correlator \eqref{correlator}. On the left side only the vacuum descendants contribute, while on the right side the descendants of the winding modes contribute. Due to crossing symmetry the non-tame functions cancel in this example. }
    \label{fig:crossingsymmetry}
\end{figure}

An example were a cancellation of non-tame behaviour within finitely many conformal blocks arises due to an additional symmetry appears in the theory of a free boson compactified on a circle of radius $R$. As this is a free theory the correlations functions can be computed using Wick's theorem and are tame functions. But in the 4-point correlation functions there still appear non-tame conformal blocks. An explicit example is given by the correlator
\begin{equation}
\label{correlator}
    \langle \partial X(x_1) e^{\frac{n}{R}}(x_2) \partial X(x_3)  e^{-\frac{n}{R}}(x_4)\rangle\;.
\end{equation}
This correlator can be expanded into conformal blocks in two different ways shown in figure \ref{fig:crossingsymmetry}. First, one can contract the $\partial X$ operators with each other or one can first contract the $\partial X$ with an exponential operator. In the first case only the vacuum descendants contribute and one immediately obtains the expected tame result. In the second case the dimension of the exponential operator is given by $\Delta=\frac{n^2}{4R^2}$, resulting in a non-tame contribution. These contributions precisely cancel and, as expected by crossing symmetry, the result is equivalent to the first expansion. This is an example of a simple cancellation due to the symmetry of the external states, but one can imagine similar cancellations to happen in more complicated scenarios.  

In the case of finitely many states such a cancellation requires a symmetry. In the case of infinitely many primaries the summation could cancel out the non-tameness even in the absence of such a symmetry. An analogy for such behaviour is the expansion of functions in terms of some basis functions. If one expands a tame function into its infinite Fourier modes, the individual terms in the expansion appear to be non-tame. This is in contrast to having only finitely many orthogonal non-tame functions, where tameness requires a cancellation of the coefficients and hence a symmetry.

Remarkably, this implication resonates with a swampland conjecture discussed in \cite{Lust:2019zwm},
which claims that there are no parametrically large 
gaps between a finite set of operators with operator dimension $\Delta_2\le\hat \Delta$ 
and an infinite set of operators with parametrically increasing dimensions $\Delta_1$.

Up to some caveats,
 \hyperlink{Conjecture4}{Conjecture 4} implies this conjecture.
One caveat is that conjecture 4 relates to continuous dependence on the parameters, while
\cite{Lust:2019zwm} also suggests it for discrete parameters.
Another caveat is
the possibility we mentioned that the full correlation functions could be tame.
Note also that there are counterexamples involving limits in which the number of degrees of
freedom diverge (more about this in \S \ref{sec:o-minQFTs}).
For example, in $N\rightarrow\infty$ (planar) 
$d=4, N=4$ super Yang-Mills, the Konishi operator (and other operators dual to string states) have $\Delta_1 \sim(g^2N)^{1/4}$ as $g\rightarrow\infty$,
while protected operators have $\Delta_2$ fixed.
However at finite $N$, the evidence is consistent with a finite upper bound for $\Delta_1$.\footnote{
The simplest argument for this 
is if the operator is an S-duality singlet, but this is not completely clear.
See \cite{Beem:2016wfs} \S 4.3.3 for various scenarios at finite $N$. }

A related but weaker claim which is easier to argue for is the following.  Let
$\Delta_1(g)$ be the dimension of an operator $O_1$ which is suspected of diverging as we take a limit in
parameter space.  Suppose we know $\Delta_1(g=0)$, and suppose the operator mixes
with another operator $O_3$ with $\Delta_3=H>\Delta_1(g=0)$ roughly constant as a function of
$g$.  Then, the phenomenon of level repulsion in quantum mechanics implies that
as we increase $g$ through the degenerate point $\Delta_1(g)=\Delta_3$, the state
$O_1\ket{0}$ will cross over to $O_3\ket{0}$ and thus the dimension $\Delta_1(g)$
will asymptote to $H$.
This is a generic statement and can be avoided by tuning enough parameters, but in a QFT with
a large number of candidates for $O_3$ one could not avoid all of the crossings.  Thus, unless
there is a symmetry reason for $O_1$ not to mix, we expect no parametric separation just for
this reason.

The conjecture of \cite{Lust:2019zwm} was also motivated by the results of 
\cite{Douglas:2006es,Tsimpis:2012tu,Gautason:2015tig} arguing for the absence of scale separation in the dual AdS gravity background. 
If it is possible to realize parametrically large scale separation by tuning
continuous parameters,\footnote{ Note that \cite{dewolfe_type_2005}
claims to realize it by tuning discrete parameters, for which this argument does not directly apply.}
then either \hyperlink{Conjecture4}{Conjecture 4} is false, 
or else one should be able to see the needed cancellations
in the full correlation functions. It would be interesting to 
investigate this in detail. 

\subsubsection*{Convergence in conformal perturbation theory}

Another long-standing related conjecture is that
conformal perturbation theory has a non-zero radius of convergence.  We recall that 
conformal perturbation theory is defined by 
adding exactly marginal operators to the action, doing a series expansion
in the associated couplings, and renormalizing divergences.
In this way one can compute the derivatives of all orders of
a general observable with respect to the moduli.   
This procedure has been long known; two recent
detailed discussions are \cite{gaberdiel_conformal_2008} in $d=2$ and \cite{sen_first-order_2018} for general $d$.
Unlike perturbation theory for general interactions added to a free field theory, 
it is generally thought that conformal perturbation theory has a non-zero radius of convergence, and thus 
general observables are analytic in the marginal couplings.\footnote{
Making this precise is subtle, because to vary the couplings one must 
renormalize the operators appearing in a correlation function, resolve mixings of degenerate operators, and
choose contact terms to respect symmetries.
The claim is that all this can be done in a way which respects analyticity.}
The folklore argument is that such perturbations do not
change the large field behavior of the action.
It is true in exactly solvable models: varying the radius of a free boson, and
marginal sine-Gordon or current-current interactions.
%And because the spectrum of the radial Hamiltonian is discrete (possibly with finite degeneracies)
%one can also apply the theory of perturbations of quantum mechanics discussed in \S \ref{sec:partition}.

If all CFT observables are definable in the o-minimal structures such as $\bbR_{\rm an,exp}$ or $\mathbb{R}_{\mathcal{G},{\text{exp}}}$ then the claim of non-zero radius 
of convergence follows from the analytic cell decomposition discussed in \S \ref{sec:o-minimalstructures}. 
They are analytic away from a finite set of lower-dimensional regions, e.g.~where they might be singular, and thus have a non-zero radius of convergence.  
Now as we discussed in \S\ref{sec:o-minimalstructures}, the analytic cell decomposition property does not hold in all o-minimal structures.  But these 
comments and the analyticity of correlation functions suggests that it should hold in $\bbR_{\text{CFT}d}$.

\section{Tameness constraints on the space of CFTs} \label{Tameness-cT}

A crucial requirement in \S \ref{sec:tameobservables} was to consider tame 
sets of quantum field theories $\cT$. This is necessary in order 
to make sense of any tameness statement about observables of these 
theories when varying them as a function of the parameters. Selecting such 
a tame set is non-trivial and it is desirable to have clear first-order 
principles to identify such $\cT$ and the associated o-minimal structure 
$\bbR_{\cT,\cS}^{\rm def}$. For effective field theories this was made concrete in 
Conjecture 2, where it was stated that the set of effective theories 
compatible with quantum gravity and valid below a fixed cutoff scale is a tame set. In this section we ask if there are other natural criteria that 
select tame sets of QFTs. We first list some of the challenges that go into defining such tame $\cT$s, and give potential ways how they can be addressed.
For CFTs we will then list a concrete set of conditions on the CFTs that 
make $\cT$ tame. Our ideas are summarized in our final conjecture:

\noindent
\textbf{\linkdest{Conjecture5}Conjecture 5 - Tameness of spaces of CFTs.}
\begin{itemize}
    \item[(a)] \textit{Let $\cT_{{\rm CFT}2}$ be the space of parameterized two-dimensional
unitary conformal field theories with central charges smaller than some fixed $\hat c>0$ 
and lowest nonzero operator dimension larger than a fixed $\hat \Delta>0$. Then $\cT_{{\rm CFT}2}$ is definable in some o-minimal structure. }
\item[(b)] \textit{Let $\cT_{{\rm CFT}d}$ with $d>2$ be the space of parameterized $d$-dimensional
unitary conformal field theories obtained by {\rm (1)} universally bounding
an appropriate measure for the degrees of freedom (e.g.~$a$ in $d=4$, or the free energy in $d=3$); 
{\rm (2)} identifying pairs of theories which are related by gauging a discrete symmetry.
Then $\cT_{{\rm CFT}d}$ is definable in some o-minimal structure.} 
\end{itemize}

We have split this conjecture into two parts, distinguishing on dimensionality. As we will explain below, we are confident that part (a) is valid due to the connection of two-dimensional CFTs to geometry. In fact, part (a) can be viewed a generalization of a conjecture made in \cite{Kontsevich:2000yf,Acharya:2006zw}. In dimension larger than two, new phenomena occur and (b) might require further refinement. In particular, we have included in (b) an equivalence relation for theories theories differing by a gauged discrete symmetry. It is likely possible to weaken this condition and it would be desirable to return to the formulation of this conjecture in the future.    

\subsection{Defining o-minimal structures associated sets of QFTs} \label{sec:o-minQFTs}

In the following we collect some general observations relevant in ensuring that $\cT$ is a definable set in an o-minimal  structure $\bbR_{\cT,\cS}^{\rm def}$. We begin with some of the challenges one has to face and then sketch our general ideas how to address them. 

\subsubsection*{Challenges for defining a tame theory space}

To get some appreciation of the difficulties that arise when trying to 
identify tame sets of QFTs, let us list some ways one would, at least 
naively, violate tameness. We do that by considering a Lagrangian QFT 
with a scalar field space $M$. The theories might also have 
a scalar potential $V:M \rightarrow \bbR$, and other coupling functions defined on $M$. Even after imposing consistency conditions on $M$ and the coupling functions, there are the following challenges to address:
\begin{itemize}
    \item There are infinitely many topological classes of manifolds in which $M$ could reside.  
    \item $M$ is allowed to have certain singularities, 
    so there are even more classes.  The allowed singularities depend on $d$ and the type of QFT.
    One class which is always allowed is an orbifold singularity produced by quotienting $M$ by a discrete
    subgroup $G$ of its isometry group.  
    Such a gauging of a discrete symmetry group $G$ was already discussed in part I, \S 4.2.  Now, if $M$ has continuous
    abelian isometries such as $U(1)$, the infinite series of subgroups $\bbZ_p\subset U(1)$ gives
    rise to an infinite set of quotients $M/\bbZ_p$ and an infinite set of distinct QFTs.
    \item Given an $M$, there are often infinitely many topologically distinct fiber bundles over $M$,
    even if we fix the fiber dimension.  For example, $U(1)$ bundles over $S^2$ are classified by the first
    Chern class, an arbitrary integer.  Physically this is the magnetic flux and this choice produces an infinite 
    sequence of quantum mechanical systems.
    \item Some couplings are quantized for other topological
    reasons, such as Chern-Simons couplings in $d=3$. These could also produce
    infinite sequences.
    \item There is an infinite set of topological field theories (TFTs), with zero Hamiltonian.  
    Given a QFT $T$, one can obtain an infinite set of theories by tensoring it with these TFTs.
    \item There could be non-tame continuous parameter spaces of physical interest,
    such as those used in the exact RG discussed in part I.
\end{itemize}
While many of these issues can be addressed in EFTs compatible with quantum gravity (see, e.g.~\cite{Palti:2019pca,vanBeest:2021lhn}) and thus further motivate \hyperlink{Conjecture2}{Conjecture 2}, we next want to develop ideas to more directly impose conditions such that 
$\cT$ is tame. 

\subsubsection*{Bounding the number of degrees of freedom} 

Let us consider the set $\widehat \cT_{{\rm CFT}d}$ of all consistent $d$-dimensional CFTs.  
While we do not know much about it, it is very unlikely
that it is tame.  Given a theory $T$, we can define the tensor product $T \otimes T$ by tensoring the Hilbert spaces and adding the Hamiltonians of the two copies, and we can define an $n$-fold tensor product $T^{\otimes n}$ analogously.  This gives us many
discrete infinite series of CFTs, and there is no reason to believe they all sit in some larger connected structure.

This particular infinity can be dealt with by upper bounding a measure of the number of degrees of freedom which
is positive for nontrivial theories and additive under tensor product, call it $F(T)$.
For two-dimensional CFTs we can take $F$ to be the central charge $c$, while in four-dimensional theories the $a$ function should be used.  Then
\beq \label{def-cT}
   \cT_{{\rm CFT}d,C} \equiv \{ T \in \widehat \cT_{{\rm CFT}d} \,|\, F(T)\le C \}\ , 
\eeq
where $C$ is a fixed positive number.  We could make a similar definition for QFTs, taking for $F(T)$
the UV limit of the central charges.

Using these sets of theories, define the structures $\bbR_{{\rm CFT}d,C}$.  Clearly increasing the bound enlarges
the set of theories,  $\cT_{{\rm CFT}d,C}\subset\cT_{{\rm CFT}d,C'}$ for $C < C'$.
However this does not imply that it enlarges the corresponding structures; it could be that the correlation functions
for the additional theories can be expressed using the same set of functions.  We only know that
$\bbR_{{\rm CFT}d,C}\subseteq\bbR_{{\rm CFT}d,C'}$ for $C< C'$.

One can entertain various hypotheses about the limiting behavior as $C\rightarrow \infty$.
The simplest is that there exists some finite $C_{\rm threshold}$ at which the limit is attained, 
\beq
\bbR_{{\rm CFT}d,C} = \bbR_{{\rm CFT}d,C'} \qquad\forall C,C'\ge C_{\rm threshold}\ .
\eeq
Not having any reason to believe the contrary, we conjecture that this is the case, and
let $\bbR_{{\rm CFT}d}$ denote this limiting structure.  The upper bound is still present, but its precise
value (at or above $C_{\rm threshold}$) does not matter.

\subsection{Tameness conditions in various dimensions}

We now want to more carefully discuss the conditions one has to impose such 
that $\cT$ could be definable in an 
o-minimal structure. A necessary condition is that $\cT$ must have finitely many connected components. This immediately 
raises several problems which we want to spell out and address in the following.

\subsubsection*{Quantum mechanics: $\cT_{\text{QFT}1}$}

Let us consider the quantum mechanics of a particle moving on a manifold $M$
in a potential $V:M\rightarrow\bbR$ and possibly with gauge fields, spin and
other couplings.  Here
there are no constraints from renormalization and thus this data naturally lives in infinite
dimensional function spaces.  We expect to require more conditions on it to get tameness.
We do not have a precise conjecture, but let us discuss some candidate conditions.

As a basic requirement on the candidate QFTs in $\cT$, we will insist that their radial Hamiltonians have discrete non-negative spectrum (for every $r$).\footnote{
There are important CFTs with continuous spectrum such as $d=2$ Liouville theory,
but we leave this for later work.}
This implies (by the definition of discrete spectrum) that eigenvalues can
only have finite degeneracy.  Furthermore, since the spectrum is bounded below,
it implies that for any $\Delta_{\rm max}\in\bbR$
there can only be finitely many states with $\Delta\le\Delta_{\rm max}$.\footnote{
Otherwise, there will be an limit point (Bolzano-Weierstrass) and the spectrum
will not be discrete.}

One could get a stronger requirement by asking that every theory in the closure
of $\cT$ has discrete spectrum.  In other words, taking the limit of a sequence
of theories does not produce a dense spectrum.  This type of limit behavior
is common, it appears for example in decompactifications.

A discrete spectrum can be guaranteed by simple conditions: for example require $V$ to be bounded from below and that for any $E$, the subset $\{x\in M|V(x)\le E\}$ has finite volume \cite{simon2008schrodinger}.  
Another possible condition (which we will definitely want in $d\ge 2$) is
a gap between the first two eigenvalues.  This is harder to enforce, but there are many
works giving sufficient conditions \cite{singer1985estimate,ling2010bounds}.\footnote{
Note also the very interesting work \cite{Cubitt2015UndecidabilityOT,Cubitt:2015lta} which shows that for spin chains, the presence of a gap can be undecidable!  These are limits of 0d systems with infinitely many
fields, so are excluded by our definitions.
}

In part I we cited work on exact WKB, which hypothetically defines a class of
quantum periods.  If these functions define a tame structure, then starting from a tame
$\cT$ we would know which observables are tame.  What conditions guarantee that
$\cT$ is tame?  A very strong conjecture would be that for {\bf any} family with
finitely many parameters (real or integer), 
any closed subset $\cT$ for which each $T\in\cT$ has discrete spectrum is tame.
This seems in the right direction but not sufficient.
Let us consider a possible counterexample with an integer parameter, namely a particle
on $S^2$ coupled to a magnetic field of flux $n$ ({\it i.e.}, first Chern class $n$).
This is not a tame set but as $|n|\rightarrow\infty$ the ground state could
become approximately degenerate.  This is true for constant $F$ and $V=0$ 
as this is the lowest Landau level.  More generally as $|F|\sim n$ becomes large,
the theory becomes semiclassical in the sense that
wave functions localize in regions of size $1/\sqrt{n}$ with energy roughly
the value of $V$ in this region.  Continuity of $V$ would then give us what we
want, but the problem is that one could choose $F$ to be zero near the minima of $V$.
But perhaps some condition like this could work.

A different approach is to make contact with finiteness results for geometry.  Given a theory $T$
in $d=0,1$ obtained by quantizing a particle on $M$, and given enough observables,
one can reconstruct $M$ with the metric
and other data.  This observation is a starting point for noncommutative geometry and is explained in
\cite{connes_noncommutative_1995,kontsevich2000homological} and related works.
To give some intuitions behind this, the complete set of $3$-point functions determines the
algebra of functions, which for a commutative algebra determines $M$ (at least as a measure space).
The metric can be determined from the Laplacian by starting with
$\Delta(fg)-\Delta(f)g-\Delta(g)f = 2\,\partial f \cdot \partial g$ and then defining the distance between
points $x,y$ as the maximum of $f(x)-f(y)$ over all $f$ with $||\partial f||^2\le 1$.

Now, there are known geometric conditions on $M$ which guarantee finiteness, such as those discussed in \cite{acharya_finite_2006} and which we will come back to in \S \ref{CFT-tameness}.  These could be
translated into conditions on $T$, for example the upper bound on diameter required by Cheeger's theorem \cite{cheeger1970finiteness}
plus a curvature condition is equivalent to a lower bound on the spectral gap.  
Perhaps these conditions can be generalized to cover all Lagrangians and translated back into quantum mechanical terms.

\subsubsection*{QFTs and CFTs in $d\ge 2$: $\cT_{\text{QFT}d}$ and $\cT_{\text{CFT}d}$ }

First, we require the ground state to be nondegenerate (as in many axiomatic definitions).
This excludes tensor products with TFTs. 

Next, we require the UV limit to be either weakly coupled (asymptotically free) or a nontrivial fixed point 
(a CFT).
Since we can list the weakly coupled theories but do not actually know all the CFTs, we might want to 
distinguish two classes of QFTs, a smaller one, denoted by AQFT$d$, which contains those QFTs that are asymptotically free, 
and a larger one QFT$d$ consisting of CFT$d$ and theories obtained by flows out of these.

Finally, let us recall an important physical difference between QFT in $d\le 2$ and in $d>2$.
Intuitively, the constant mode of a scalar field is quantum in $d\le 2$ and classical in $d>2$.
Thus an energy eigenstate in $d=2$ is delocalized, as in quantum mechanics.  These delocalized states
might be further decomposed into states of definite momentum (if there is translational symmetry), 
group representation matrix elements (for group symmetry), or more generally states related to
eigenfunctions of the Laplacian on $M$.  By contrast, in $d>2$ the fluctuations of a scalar fall off
with distance, and one can fix the zero mode by using sources or boundary conditions.  Thus the vacuum
decomposes into superselection sectors labeled by the scalar expectation values, which must furthermore
be critical points of the effective potential $V'_{\rm eff}(\langle\phi\rangle)=0$.  The space of these is now part of the theory space $\cT_{\text{QFT}d}$ and will be referred to as the vacuum moduli space.
Thus a complete definition of ``space of QFTs'' in $d>2$ must account for this, however movement in $\cT_{\text{QFT}d}$
breaks conformal invariance (since operator dimensions must be positive by unitarity) and we will not
get far enough in our discussion of more general QFT to need the details.

What we will use below is the geometric intuition related to this difference.  In $d\le 2$ there are QFTs
which are generalizations of manifolds (sigma models, with $V=0$), those which are generalizations of singularities
when $V(\phi)$ has isolated critical points, and hybrid theories ($V(\phi)$ is non-zero and has a higher dimensional manifold
of critical points).
In $d>2$ one only has the singularities.

\subsection{Tameness of spaces of
conformal field theories} \label{CFT-tameness}

In the following we investigate the tameness of the spaces 
of $d$-dimensional CFTs $\cT_{{\rm CFT}d}$  and holographic CFTs $\cT_{\text{AdS}_{d+1}}$. These theories have a fairly satisfactory rigorous definition in terms of its spacetime symmetry and the Osterwalder-Schrader axioms\footnote{
These axioms define the Euclidean spacetime theory and
include reflection positivity (unitarity in physics parlance) and
single-valuedness of correlation functions.  We will also assume the state-operator correspondence
of radial quantization and the existence of an identity operator.
}
(see \cite{kravchuk_distributions_2021} for a recent discussion). The 
challenge will be to restrict the set of CFTs in such a way 
that there exists an o-minimal structure $R_{\cT,\cS}^{\rm def}$ in which 
they are definable. The aim of this section will thus be to further motivate 
the assumptions of \hyperlink{Conjecture5}{Conjecture 5}.

\subsubsection*{Tameness in $d=2$ CFTs}

As a first case we discuss two-dimensional CFTs. These theories have been studied very extensively and rather well understood. Furthermore, there are previous conjectures about the space of these theories \cite{Perlmutter:2020buo} and we will comment on their relation to our \hyperlink{Conjecture5}{Conjecture 5} (a).  

To begin with let us motivate the constraints in \hyperlink{Conjecture5}{Conjecture 5} (a) that we 
conjecture ensure that $\cT_{\rm CFT2}$ is tame. 
First, we are discussing sets of CFT's with an upper bound $\hat c$ on the central charge,
as discussed in \S \ref{sec:o-minQFTs}.  But even with this condition, there are
evident counterexamples to tameness, such as the set of $c<1$ unitary minimal models.
These have $c=1-6/(p+1)(p+2)$ and are clearly not connected within the space of CFTs.
Another example of a $d=2$ infinite series is the $SU(2)_k$ WZW models,
with $c=3-6/(k+2)$.

These two infinite series can be cut off by imposing a positive lower bound on the dimensions of nontrivial operators.  
For the minimal models, the lowest nonzero operator dimension is $\Delta_1=3/4(p+1)(p+2)$ (for $p\ge 2$), and
for $SU(2)_k$ WZW it is $\Delta_1=3/4(k+2)$.  Thus this works for any positive lower bound.

Why should this particular bound lead to finiteness more generally?  To see this, first note that both
of our examples are associated with limits in which the diameter of the target space becomes large.
The infinite series of WZW models arises from a $\bbZ$-valued topological choice (the
quantization of the WZW coefficient $k$ is topological).  In fact the metric at the conformal fixed point
is proportional to the WZW term, so this is a large target space limit.  
As for the minimal models,   they can be
identified as the multicritical points of a theory of a real scalar field $\phi$, in other words one
takes a polynomial potential $V_p$ of degree $2p$ with lower order coefficients tuned to achieve the
largest central charge at the IR fixed point.  Intuitively we can think of the potential as confining
the scalar field to an interval of length $L$ (defined using the metric on field space).
The most relevant operator is then $\phi$.  Now, as $c\rightarrow 1$ the theory becomes semiclassical
and the dimension of $\phi$ approaches that of a free boson, $\Delta \sim 1/L^2$.  Again we need
$L \sim p \rightarrow \infty$ to reproduce the series of dimensions.  We conjecture that this relation
is more general, so any infinite series will be associated to a limit of large target space diameter.

This turns the finiteness question into finiteness for a sequence of target spaces with bounded dimension.
As discussed in \cite{acharya_finite_2006}, this finiteness follows from Cheeger's theorem under the assumption
of upper bounded curvature (so, perturbation theory is good) and lower bounded diameter.  Now the leading perturbative
estimate for the lowest nonzero operator dimension is the lowest nonzero eigenvalue of the scalar Laplacian $\lambda_1$, and this
is related to the diameter of the target space as $\lambda_1\sim 1/\mbox{diameter}^2$.  Thus, if this operator is in the spectrum,
any positive lower bound $\Delta_1\ge \Delta_{min}>0$ will work.\footnote{ This bound would fail if the corresponding
operator is not in the spectrum, for example if the theory were GSO projected.  However, the axioms we cited earlier
do not allow this case. } Note that imposing this bound makes the conjecture weaker. For example, it then does not directly imply that there is a finite number of Calabi-Yau threefolds, but it does imply that
if there is an infinite series of these then the gap $\Delta\rightarrow 0$.  Since the Ricci flatness condition is scale
invariant, one cannot say that the diameters go to infinity (after all these are controlled by a K\"ahler modulus).  But
presumably if we consider an infinite sequence with bounded diameter, the maximal Riemann curvature would grow without limit.

There is also a relation to the distance conjectures of Refs. \cite{Baume:2020dqd,perlmutter_cft_2020}
(and of earlier work cited there).  Granting these conjectures, the lower bound on operator dimensions
excludes any infinite distance limit along a conformal manifold (varying marginal couplings).  This 
resonates with our previous discussion and is
helpful as such infinite distance limits could be a source of non-tame behavior.  The conjectures 
formulated in Refs. \cite{Baume:2020dqd,perlmutter_cft_2020} do not directly address discrete series of theories,
but perhaps this could be done with a more general definition of distance as in \cite{douglas_spaces_2010}.

Just as we conjectured in \S \ref{sec:o-minQFTs} that the structure $\bbR_{{\rm CFT}2}$ does not depend
on the particular (sufficiently large) upper bound on the central charge, so too we conjecture
that we get the same structure $\bbR_{{\rm CFT}2}$ for any (sufficiently small)
lower bound $\Delta_{\rm min}>0$.  A physical argument is that the large volume limit is weakly coupled
(from the $2$-dimensional point of view) and shares the simplicity of weak coupling limits. While we have seen in part I that 
a possible choice for $\bbR_{{\rm CFT}2}$ is $\bbR_{\rm an,exp}$, 
we have seen evidence in \S \ref{sec:tameCFTobservables} that one needs to go beyond this structure. We leave this as an open question for future research.

\subsubsection*{Boundary CFT in $d=2$ } 

One can make a parallel discussion of boundary conditions in a fixed 2d CFT. Now it is natural to put an upper bound on the boundary entropy.
This directly controls the spacetime volume and also
solves the problem mentioned in \S \ref{sec:o-minQFTs} that there are
an infinite number of topological classes for the gauge field, as the
boundary entropy grows with the field strength (in D-brane language it
is the world-volume tension, given by the Born-Infeld action).

\subsubsection*{Issues in $d\ge 3$ -- discrete gauge symmetry } 

Let us now turn to \hyperlink{Conjecture5}{Conjecture 5} (b). In higher-dimensional CFTs one can make a very similar conjecture to the $d=2$ case, but these spaces of CFTs are less well understood.
Let us discuss some examples which illustrate new points which arise.

First, consider a theory with a $U(1)$ global symmetry group $G$.  For definiteness we will consider two examples,
$M_1=S^1$ with the shift symmetry $\phi\rightarrow\phi+\theta$, and $M_2=\bbC$ with the
rotational symmetry $\phi\rightarrow e^{i\theta}\phi$.  Now $\bbZ_p\subset G$ and thus we can quotient
(discrete gauge) $M$ by $\bbZ_p$.  What about these infinite series?

Of course the quotients $S^1/\bbZ_p$ are nothing new -- they are circles of reduced radius $R\rightarrow R/p$.
Thus they form part of a connected moduli space and lead to no problems.  But since the new target spaces are smaller,
this infinite series does not seem to respect our intuition that infinite series are associated with growing 
target spaces and a vanishing gap. The resolution of this in $d=2$ is familiar.  It is that there are winding states
with operator dimension $R^2/p^2$ (in appropriate units) and the gap does vanish.  Even if we did not have this geometric
intuition, we would discover this once we added twist sectors to obtain a modular invariant theory.  Very generally,
discrete gauge symmetry leads to sectors with periodic boundary conditions twisted by a gauge transformation, and in
this case these are the winding states.

What happens to this argument in $d>2$ ?  The direct analog of the $d=2$ twist sectors are present, but only
for a state defined on a noncontractable spatial slice, such as $S^1 \times X$.  These are associated with
operators of codimension $2$, and are nonlocal in $d>2$, so do not enter directly into the definition of the gap.
While the $M_1=S^1/\bbZ_p$ example is still no problem, now the $M_2=\bbC/\bbZ_p$ series of theories seems to provide
a infinite series without a vanishing gap (the noncompactness of $M_2$ is not a problem in $d>2$).

If this is so, then to save the conjecture we need to place another condition which takes care of the infinite series that arise from these theories. We have done this in \hyperlink{Conjecture5}{Conjecture 5} (b).  
Alternatively, one might put a condition the related $d=2$ theory obtained by
compactification on $S^1$ and in which there will also be conventional point-like
operators of low dimension.  A problem with this is that the compactified theory is only
a CFT in the IR so there is an implicit use of the RG in this proposal.
This would make it hard to use and does not sound right.

A promising direction is to replace the equivalence relation in \hyperlink{Conjecture5}{Conjecture 5} (b), to include a condition on extended states. More precisely, we can interpret the twist operators as creating extended states,
not present in the usual perturbative discussion,
and study the dimensions (radial energies) of these states.  It could be that these states actually have small energy $\Delta=(d-2)/2+\cO(1/p)$, in which case a gap condition
would again suffice to remove non-tame infinite series of theories.

\subsubsection*{Issues in $d= 3$ -- Chern-Simons terms } 

Another infinite series specific to $d=3$ is the series of integral values of
a Chern-Simons coupling.  Such a set of theories could still be tame under
our definitions if placing an upper bound on the free energy on $S^3$ removes
all but a finite number.  This question does not seem to have been explicitly
addressed in the literature and we study it in examples in Appendix A.

We find that it is indeed the case, for two reasons.  First, while it seems
evident that $F\sim N^2$ could bound the allowed values of $N$, one might worry
about infinite series in $k$.   But at fixed $N$, the
$U(1)_k$ and $SU(N)_k$ pure CS theories have $F \sim \log k$.  More
generally, we will see that the leading behavior is $F \sim \min(N,k)^2\log(N+k)$.

Second, once we know that the pure CS theories are bounded,
then if the additional contribution of matter to $F$ is positive, adding
matter can only reduce the number of theories satisfying a given bound.
Now according to the F-theorem, an ungauged matter theory 
must have a positive free energy.  This is less obvious for the interacting
theory but it is the case for the conjectured exact results \cite{Giombi:2017txg}.

\subsubsection*{Geometric picture in $d\ge 3$ } 

As we discussed earlier, the geometric picture relevant to $d>2$ QFT is that
of singularity theory.  Often the vacuum moduli space is a cone over an space
$X$ and thus one can look at geometry of $X$.  This is particularly true in
AdS/CFT where the vacuum moduli space is $X^N/S_N$ or a deformation of this.

If we restrict attention to theories with geometric duals, we can make
contact with finiteness results for these.  There are general arguments
in Ref. \cite{acharya_finite_2006} which relate to a number of examples
we discussed.  One basic phenomenon is that reducing the volume of $X$
increases the AdS radius and thus the number of degrees of freedom of
the dual CFT.  This removes infinite series produced by quotienting
by discrete symmetries.

Consider the supersymmetric orbifolds of maximal super Yang-Mills,
for example the $\bbC^2/\bbZ_p$ orbifolds
studied in \cite{Kachru19984DCF} and many subsequent works.  
These are quiver theories with 8 supercharges and
$O(N^2 p)$ degrees of freedom, manifestly
at weak coupling, and also in strong coupling by AdS/CFT.

As another example, many CFT3's with geometric duals 
can be understood as theories of M2-branes probing singularities.
For example the ABJM theories \cite{aharony_n_2008} arise as probes
of $\bbC^4/\bbZ_k$ and are dual to $AdS_4\times S^7/\bbZ_k$.
The parameter $k$ becomes the Chern-Simons level
$k$ in the dual, and now the effect of the quotient on the central charge
is another example of the phenomena we discussed in the previous subsection.

In a different direction, one could go on to conjecture that for $d>2$, 
the vacuum moduli space $\cS$ of each such CFT is definable in an o-minimal structure.
Since general expectation values break conformal symmetry,
this brings us to our next topic.

\subsection{Arguments from microscopic definitions}

Ideally one would prove tameness statements by starting with a
space $\hat{\mathcal{T}}$ of ``pre-theories'' with a simple definition,
embedding the theory space  $\mathcal{T}$
in $\hat{\mathcal{T}}$,
and then showing that $\hat{\mathcal{T}}$ and the embedding are tame.  

To illustrate this
approach, let us outline a strategy for proving the tameness of a space of CFTs,
based on conjecture 2 of \cite{douglas_spaces_2010}.
This conjecture states that all $d=2$ CFTs with central charge less than a given upper bound 
$c$ can be obtained as RG flows from a sum of free CFTs and WZW models with at most $N(c)$ factors.
A generalization to $d\ge 2$ is to take $\hat{\mathcal{T}}$ to be a space of asymptotically free theories which can be defined 
in the UV using resummed perturbation theory, in terms of a fixed number $N$ of UV fields.
This space is parameterized by the couplings $g$ in the UV
Lagrangian, so it is plausible that we could define a tame region 
$\mathcal {T}\subset\hat{\mathcal{T}}$ as the region (schematically)
$||g||< 1$.  We would then need to prove that every CFT with $c\le c(N)$ can be obtained by RG flow
from $\mathcal {T}$, and that this RG flow map is tame.

Another strategy might be based on the conformal bootstrap.  Thus, we hypothesize (as in 
conjecture 1 of \cite{douglas_spaces_2010})
that given $c$, there is some $h(c)$ such that the spectrum and three point functions for all operators of dimension
up to $h(c)$ uniquely determines the theory; this data then defines a point in $\hat {\mathcal {T}}$.
Further assuming an upper bound $N(c)$ on the number of these operators, we might somehow show that $\hat {\mathcal {T}}$
is tame; the bootstrap inequalities then (being finite in number) mark out a tame subset of true CFT's
$\mathcal {T}\subset\hat{\mathcal{T}}$.

\section{Conclusions} 

In this work we formulated a procedure by which, given a family (or set) of QFTs $\cT$,
and ``computations'' of a definite set of their observables, namely partition and correlation functions,
one can derive a structure $\bbR_\cT$ (as defined in \S \ref{sec:structureQFT}).
This new perspective uses basic ideas from first order logic in which a structure provides a language to formulate logical statements, which in our case are statements about the family of physical theories and their observables. 
The structure $\bbR_\cT$ can admit various general mathematical properties. Central to this work was a property called
o-minimality or tameness, which 
imposes an intricate finiteness condition on the structure. 
In particular, observables in o-minimal structures cannot have too wild behavior and very general questions about these have finite answers.
Showing that the structure defined by a set of QFTs is o-minimal is a very general way to answer
finiteness questions studied in physics -- of number of components of their parameter space and space of vacua, 
of the number of critical points, or about other structure of observables.

The quote marks around ``computations'' above is meant to indicate that one does not need exact results
to derive a QFT structure, only general statements about what types of correlation functions arise
and what structures they live in.  In some cases it suffices just to know the domain of definition of the
functions (in other words the location of their singularities) to identify a structure. 
The mathematical literature contains many examples for interesting classes of structures, statements that large classes of functions
live in particular structures, and proofs that certain structures are o-minimal.  The framework is
compositional, so that if one knows that subsets of the correlation functions each live in particular
structures one can use this information to constrain or even derive the union structure.
For all of these reasons, and especially because many challenging problems in mathematics have been solved using
these techniques,
we believe the problem of deriving the structures associated to particular classes of QFTs will be far
more tractable than exactly solving the theories or any of the other existing approaches to studying ``theory spaces''.

Our definition of a structure consisted of two steps. First, we carefully selected theory spaces and sets of Euclidean spacetimes which can be described using tame sets and functions of some o-minimal structure $\bbR_{\cT,\cS}^{\rm def}$. Second, we extended this structure by adding observables of these tame theories on the tame spacetimes and claimed that the o-minimality property is preserved for the resulting larger structure. In one well known approach to Lagrangian QFTs, the observables are certain path integrals summing up intermediate configurations. Evaluating such integrals is an art and often requires intricate physical reasoning. It is unlikely that all necessary concepts to perform these integrals are definable in an o-minimal structure. Our claim is different; namely, we claim that in  certain settings we expect that the physical answers can be added to an o-minimal structure $\bbR_{\cT,\cS}^{\rm def}$ resulting in a (possibly larger) o-minimal structure $\bbR_{\cT,\cS}$. Note that this claim can be viewed as a  generalization of a mathematics conjecture \cite{vandenDriesConj,KAISER20121903} (see \hyperlink{Conjecture1}{Conjecture 1}) that adding ordinary integrals to an o-minimal structure preserves o-minimality.       

As already stressed in part I and in \cite{Grimm:2021vpn} we expect that tameness depends on the UV properties of the considered theories. 
Conformal field theories are a natural setup in which we expect tameness to be generally realized. This has lead us to propose two concrete conjectures. 
The first, \hyperlink{Conjecture4}{Conjecture 4}, asserts that observables of tame sets of CFTs are tame, while the second, \hyperlink{Conjecture5}{Conjecture 5},
claims that sets of $d$-dimensional CFTs satisfying
certain precise conditions (most notably a bound on the number of degrees of freedom) are tame. We have motivated the tameness of CFT observables by examining the spacial functions 
arising in specific CFT examples 
and in the general conformal partial wave expansion of correlators. For the tameness of sets of CFTs, we highlighted some particularly challenging examples, like 3d Chern-Simons 
theories, and showed that infinite discrete choices are avoided. While our evidence is non-trivial, it is desirable to more systematically investigate examples or even prove the conjectures 
using the axiomatic approach to CFTs.  
Whether or not there exists a maximal o-minimal structure $\bbR_{{\rm CFT}d}$ that allows to define all sets of CFTs specified in \hyperlink{Conjecture5}{Conjecture 5} and their observables remains as another interesting open question for future work. While this possibility is appealing, the general theory of o-minimal structures also allows for complicated webs of structures that are not unified in a maximal one.  

In our study of CFTs we have highlighted that the proposed tameness can lead to interesting consequences when combined with known 
properties of observables. Our reasoning was that if we know the functional dependence of CFT observables, say on the conformal dimensions of operators, and can argue that it is non-tame when freely probing the domains of those functions, then we must be missing constraints that ensure tameness. For example, we have seen that under certain assumptions the parametric separation of two operator dimensions can lead to non-tame behaviour and therefore should be prohibited. Arguments of this type can be the basis of many future investigations and will allow for establishing links to previously suggested swampland conjectures. While tameness at first might not seem very powerful, we expect it to have a rich set of implications when combined with other, possibly known, properties of a theory. 

We have also considered effective theories that can be consistently coupled to quantum gravity in \hyperlink{Conjecture2}{Conjecture 2} and \hyperlink{Conjecture3}{Conjecture 3}. Again we have made claims about the tameness both of the theory space and the space of observables. In this case, the unifying o-minimal structure for a $d$-dimensional theory was denoted by $\bbR_{{\rm EFT}d}$ and, at least in its definition, depends on the cutoff energy below which the considered theories in $\cT_{{\rm EFT}d}$ are valid. As it is common to all conjectures dealing with consistency with quantum gravity, also  \hyperlink{Conjecture2}{Conjectures 2} and \hyperlink{Conjecture3}{3} are highly speculative and eventual proofs hinge on gaining a better understanding of quantum gravity itself. Nevertheless, the conjecture can be falsified if effective theories derived, for example, within string theory violate tameness. While no such examples are known to us, it is desirable to extend the search and specifically look for possibly non-tame observables within string theory. Clearly, it is also an open question whether there exists a maximal o-minimal structure capturing all effective theories that arise from string theory.

Let us close by mentioning some of the possible further applications of our construction. As stressed in the conclusion of part I, we expect that some of the recent mathematical theorems using o-minimal structures will find direct applications to physical systems. In particular, some of the recent results, see e.g.~\cite{BT,GaoKlingler,Chiu,BTnew}, can be read as linking relations among observables of a theory to symmetries of its formulation. Focusing on the theory space, one can aim to extend the finiteness results of \cite{Bakker:2021uqw}
to larger classes of QFTs, EFTs, string compactifications and quantum gravity theories.  Another is to
provide an invariant which can be used to distinguish sets of QFTs, by showing that they give rise to
different o-minimal structures.  A related idea is to show that a proposed computational technique would not
suffice to solve a given class of QFTs, by arguing that all the functions which can be produced by this
technique generate a structure which does not define all the correlation functions of the QFTs.

\subsubsection*{Acknowledgements}

We would like to thank Ofer Aharony, Ben Bakker, Matthias Gaberdiel, Matt Kerr, Tobias Kaiser, Bruno Klingler, Maxim Kontsevich, Eran Palti, Julio Parra-Martinez, Erik Plauschinn, Volker Schomerus,  
Zohar Komargodski, Stefan Vandoren, and Mick van Vliet for useful discussions and comments. The research of TG and LS is supported, in part, by the Dutch Research Council (NWO) via a
Start-Up grant and a Vici grant.

\newpage
\appendix

\section{Cell decomposition in tame geometry} \label{app-cells}

In this appendix we give a brief account of the cell decomposition theorem following \cite{VdDbook}. 
The following holds for any o-minimal structure $\cS$ and we will talk about definability without always referring to $\cS$.

A foundational result needed for the cell decomposition theorem is the monotonicity theorem. Consider a definable function $f:(a,b) \rightarrow \bbR$, where $a,b$ can also be $\pm \infty$. The theorem states that $(a,b)$ admits a finite 
subdivision, i.e.~a split
\begin{equation}\label{split-int}
    a=:a_0 < a_1 < ...< a_{m-1} < a_m:=b\ ,
\end{equation} 
such that $f$ is either constant, or strictly monotonic and continuous on the open intervals $(a_k,a_{k+1})$.
Accordingly, $f$ can only have finitely many discontinuities. A stronger version of this theorem ensures that $f$ is $C^p$ on open intervals in a finite decomposition. Note that one should think of the division of $(a,b)$ as being a split into finitely many `cells', namely points $\{a_k\}_{k=1,...,m-1}$ and open intervals $\{(a_k,a_{k+1}) \}_{k=0,..,m-1}$. The monotonicity theorem constrains the behaviour of $f$ on these cells.

We next introduce the cell decomposition of $\bbR^n$. 
For $\bbR=(-\infty,\infty)$ we want to describe the split \eqref{split-int} in a way that generalizes to higher dimensions. Starting from $\bbR^0$, a point $p$, we define a set of functions $f_m:\bbR^0 \rightarrow \bbR$ such that $f_m(p)=a_m$. The cell decomposition of $\bbR= \bbR^0\times \bbR$ is then given by cells $\{(p,a_m) \}$, i.e.~the graphs of the functions $f_m$, and cells that are the bands $\{(p,y): y\in (f_m(p),f_{m+1}(p))\}$. The functions $f$ are trivially definable and continuous. 
This construction can now be generalized iteratively giving a definition of a \textit{cell decomposition of $\bbR^n$}: 
\begin{itemize}
 \item For $\bbR^0$ there is a unique cell, the point $p$. 
 \item For $n>0$, we write $\bbR^n = \bbR^{n-1} \times \bbR$. Given the cell decomposition $\{ \cC_\alpha\}$ for $\bbR^{n-1}$, one introduces for each cell $\cC_\alpha$ an integer $m_\alpha>0$ and a set of definable continuous functions $f^{(\alpha)}_k: \cC_\alpha \rightarrow \bbR$ for $0< k < m_\alpha$ 
such that 
\beq
   -\infty =: f^{(\alpha)}_0  < f^{(\alpha)}_1 < \ldots < f^{(\alpha)}_{m_\alpha -1} < f^{(\alpha)}_{m_{\alpha}} := \infty\ ,  
\eeq
is true on all of $\cC_\alpha$. Having such a set of functions the cells in $\bbR^n$ are: \\
(1) graphs of functions:
   $\{(x,f^{(\alpha)}_k(x))\subset \bbR^n:x\in \cC_\alpha \}$ for each 
$\cC_\alpha$;  \\
(2) bands between functions: 
$\{(x,y)\subset \bbR^n: x\in \cC_\alpha, y\in (f^{(\alpha)}_k(x),f^{(\alpha)}_{k+1}(x))\}$. 
\end{itemize}
Note that by construction the cell decomposition gives 
a split of $\bbR^n$ into finitely many pairwise disjoint definable subsets. By definition these cells are build from definable functions that are continuous. One can also demand that they are in $C^p$, which then leads to a $C^p$-cell decomposition.   

\begin{figure}
\begin{center}
\vspace*{.5cm}
 \includegraphics[width=0.65\textwidth]{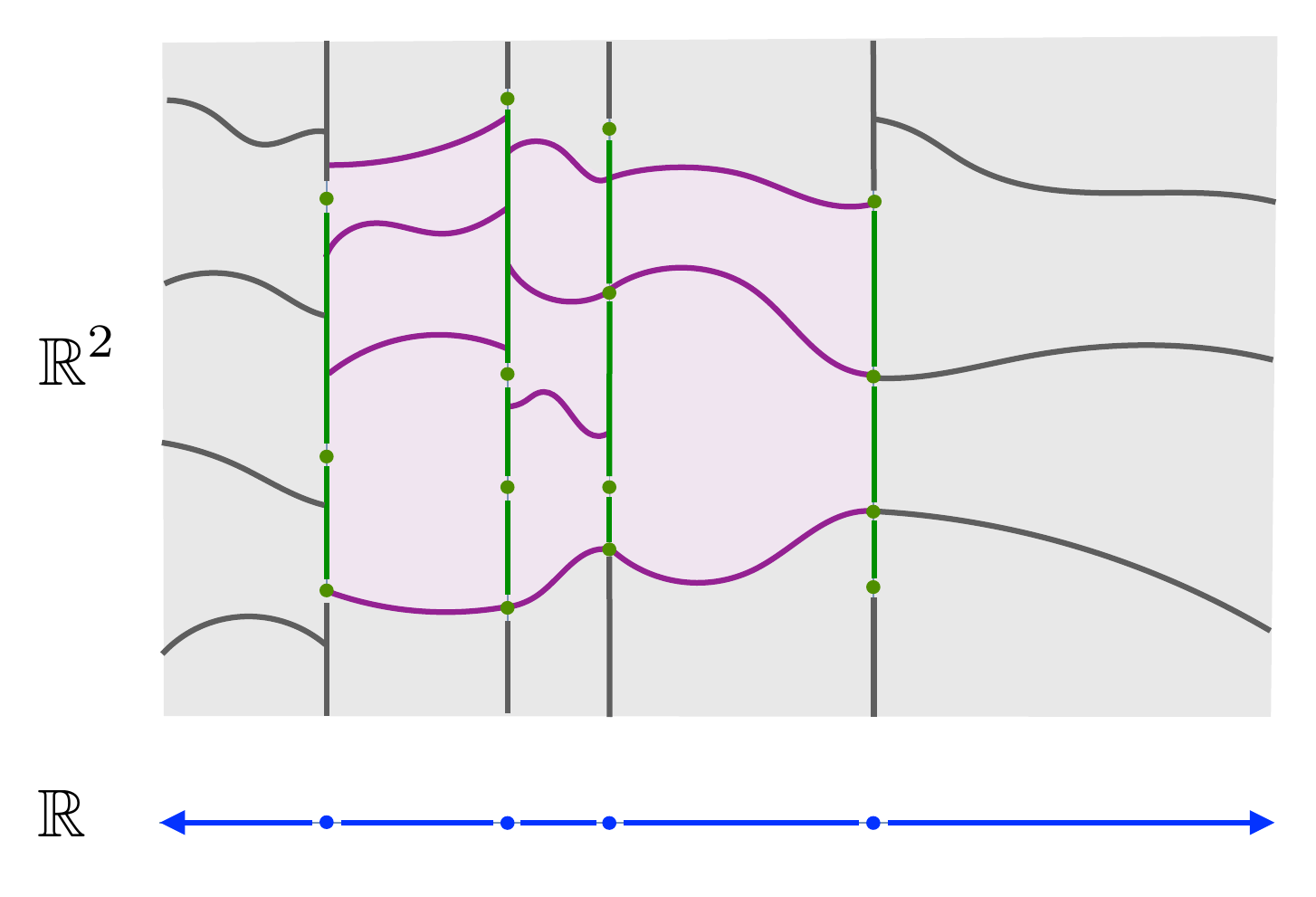} 
 \vspace*{.5cm}
\caption{The picture shows a cell decomposition of $\bbR^2$ (upper part) that is built from a cell decomposition of $\bbR$ (lower part). The new $\bbR^2$-cells
are graphs and bands of definable functions on the $\bbR$-cells. The new cells over 
open intervals in $\bbR$ are shown in purple, while the new cells over points in $\bbR$ are shown in green. 
We have coloured all cells stretching to $\pm \infty$ in grey.  \label{def-decR2}}
\end{center}
\vspace*{-.8cm}
\end{figure}

We also stress that the cell decomposition has 
special directions along which there is a simple 
projection to a low-dimensional cell decomposition. 
In figure \ref{def-decR2} a cell decomposition of $\bbR^2$ and its underlying decomposition of $\bbR$ is shown.

The \textit{cell decomposition theorem} now states that definable cells are enough to describe any definable set. The theorem states 
\begin{itemize}
    \item[(1)] For any finite collection of definable sets $A_1,...,A_k \in \bbR^n$ there exists a cell decomposition such that each $A_i$ is a finite union of cells;
    \item[(2)] For each definable function $f:A\rightarrow \bbR$, $A\subset \bbR^n$ there is a cell decomposition of $\bbR^n$, partitioning $A$ as in part (1), such that $f$ restricted to each cell is continuous.
\end{itemize}
 As stressed already in \S\ref{sec:o-minimalstructures}, this  theorem can be generalized by replacing `continuous' with being in $C^p$. In this case one uses a $C^p$-cell decomposition and infers in part (2) of the theorem that every definable function has a split into $C^p$-pieces.

The cell decomposition theorem allows to associate a dimension and an Euler characteristic to every definable set. Let $A$ be such a definable set and let $C_i$ be the finite number of cells partitioning $A$. The  
dimension $\text{dim}(A)$ of $A$ is then defined to be maximal dimension $\text{max}\{ \text{dim}(C_i)\}$ of the appearing cells.
The Euler characteristic is defined to be 
\begin{equation}
     E(A) = \sum_{C_i} (-1)^{ \text{dim}(C_i)} = n_0 - n_1 +...+(-1)^{\text{dim}(A)} k_{\text{dim}(A)} \ ,
\end{equation}
where $n_s$ is the number of $s$-dimensional cells in $\{C_i \}$. Note that $\text{dim}(A)$ and 
$E(A)$ are independent 
of the chosen cell decomposition and thus associated to the set $A$ and not the decomposition.

\section{Three dimensional Chern-Simons theories}

In this appendix we discuss the theory space of some three dimensional Chern-Simons theories with matter fields. We split the discussion into several parts, depending on the amount of supersymmetry. For generic parameters there is a non-vanishing $\beta$-function. As we are most interested in conformal field theories, we will restrict the parameter space to the conformal fixed points.
We also focus on large but finite $N$ and neglect the difference between a $U(N)$ and $SU(N)$ theory, so the level of the $U(1)$ factor does not matter. 

Our focus question will be, out of the infinite series of theories with Chern-Simons level $k$ and arbitrary matter, what is the 
universal term $F(N,k)$ in the free energy on $S^3$, and does bounding $F\le F_{max}$ give us a finite number of theories. 
By use of parity we can take $k>0$.
There is an important subtlety in the definition of $k$ -- it is shifted by one-loop effects
of the fermions, in a way which depends on the choice of UV regulator.
We refer to \cite{aharony_baryons_2016} and 
appendix D of \cite{Aharony:2018pjn} for a detailed explanation.
One must be careful to use the right conventions.

The large $N$ limiting theories are parameterized by one of $N$ or $k$ and one of the `t Hooft couplings
\begin{equation}
\lambda=\frac{N}{k} \in (0,\infty); \qquad \lambda' = \frac{N}{k+N} = \frac{\lambda}{1+\lambda} \in (0,1).
\end{equation}
These domains of definition are assuming the naive bounds $N,|k|>0$.
but as we will see below consistency can impose stronger bounds.

\subsubsection*{Pure Chern-Simons theory}

\def\tk{{\tilde k}}

Let us start with the pure CS theory with a single $U(N)$ gauge field and
CS levels $k$ and $\tk$ for the $SU(N)$ and $U(1)$ factors respectively.
The free energy $F=-\log Z$ on $S^3$ is
\begin{align} \label{eq:FpureCSa}
    F_{CS}(N,k,\tk=k+N) &= \frac{N}{2}\log(k+N)-\sum\limits_{j=1}^{N-1}(N-j)\log(2 \sin \frac{\pi j}{k+N})\\
    F_{CS}(N,k,\tk) &= F_{CS}(N,k,\tk=k+N) + \frac{1}{2} \log \frac{\tk}{k+N} \;.
\end{align}
It satisfies level-rank duality
\begin{equation}
    F_{CS}(N,k,\tk)=F_{CS}(k,N,\tk)\;.
\end{equation}
Bounding $F$ removes the $\tk\rightarrow\infty$ $U(1)$ infinite series, so let us
take $\tk=1$ (which is also the $SU(N)$ result).

We need to check that as $N\rightarrow\infty$ at fixed $\lambda$, $F(N,\lambda)\rightarrow\infty$.  {\it A priori} the
$N\rightarrow\infty$ and $k\rightarrow\infty$ limits need not commute, so
we should also check it as $k\rightarrow\infty$ at fixed $N$.  
This is easy for $N=2$, where
\begin{equation} \label{eq:CSNtwo}
        F_{CS}(N=2,k,\tk=1) = \frac{1}{2}\log(k+2)-\log(2 \sin \frac{\pi}{k+2}).
\end{equation}
This diverges for large $k$ as $(3/2)\log(k+2)$, so the $SU(2)_k$ infinite series
also runs off to large $F$.

The full $1/N$ expansion (first taking $k\rightarrow\infty$ then $N\rightarrow\infty$)
can be evaluated analytically using Barnes G-function, with the result \cite{Periwal:1993yu}
\begin{equation}
\begin{aligned}
    F_{CS,k\rightarrow\infty}(N)=&\frac{1}{2}\log(\lambda')+N^2(\frac{3}{4}-\frac{1}{2}\log(2\pi \lambda'))+\frac{B_1}{2}\log(N)\\&+\sum_{l=2}^\infty N^{2-2l}(-1)^l
\frac{B_l}{2l(2l-2)},
\end{aligned}
\end{equation}
where $B_l$ are the Bernoulli numbers. Note that the $k\rightarrow\infty$ limit corresponds for fixed $N$ to $\lambda'\rightarrow 0$. Thus the free energy diverges in this limit for any value of $N$, cutting off the tower of theories.

\begin{figure}[h!]
    \centering
    \includegraphics{"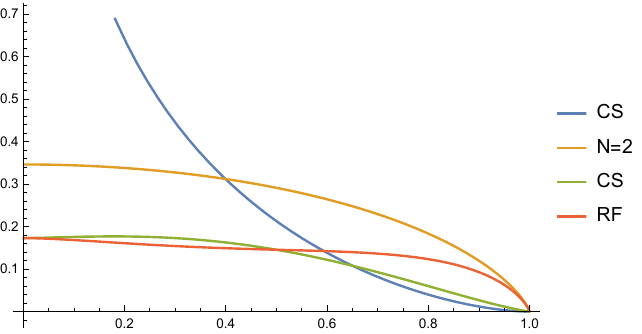"}
    \caption{Free energies $F_{CS}$, $B_{N=2}$, $B_{CB}$ and $B_{RF}$ as functions of $\lambda$ or $\lambda'$. }
    \label{fig:my_label}
\end{figure}

The large $N$ and $k$ asymptotics keeping $\lambda'$ fixed of Eq. \ref{eq:FpureCSa} are
\begin{align}\label{eq:defFCS}
F_{CS} &= N^2 F_{CS}^{(0)} + {\mathcal O}(\log N) \\
\label{eq:giombiN2F}
F_{CS}^{(0)}(\lambda') &= - \int_0^1 dx (1-x) \log 2\sin(\pi\lambda' x) \\
&= \frac{1}{8\pi^2(\lambda')^2}\left(2\zeta(3)-Li_3(e^{2 i\pi \lambda'})-Li_3(e^{-2 i\pi \lambda'})\right)
\end{align}
The function $\Re Li_3(z)$ is bounded on the unit circle and takes values between $Li_3(-1)=-3\zeta(3)/4$ and $Li_3(1)=\zeta(3)$.
As $\lambda'\rightarrow 0$ (and similarly 1), 
\begin{equation}
    F_{CS}^{(0)}(\lambda') \sim \frac{3}{4} - \frac{1}{2}\log 2\pi\lambda' + \frac{\pi^2(\lambda')^2}{72} + \ldots
\end{equation}
So it is large for small $\lambda'$.
If $F_{CS}^{(0)}(\lambda')$ had a positive minimum in $(0,1)$ then we could conclude finiteness. However, the graph in Figure \ref{fig:my_label} suggests that it does not.
Using the duality to write
\begin{equation} \label{eq:FCSlimone}
F_{CS}^{(0)}(1-z) = \frac{z^2}{(1-z)^2} F_{CS}^{(0)}(z) = z^2 \left( 1 + {\mathcal O}(z) + \ldots \right) 
\left( \frac{3}{4} - \frac{1}{2}\log 2\pi z + {\mathcal O}(z) \right) ,
\end{equation}
we see that it has a zero at $\lambda'\rightarrow 1$.
Thus we need to look at this limit more closely to see if an upper bound
on $F_{CS}$ gives us a finite number of theories.  Restoring $N$ and $k$,
\begin{equation}
    F_{CS}(N,k) = k^2
    \left( \frac{3}{4} - \frac{1}{2}\log \frac{2\pi k}{N+k} + \ldots \right)
\end{equation}
So the $k\rightarrow\infty$ limit is cutoff by the overall $k^2$ dependence,
while the $N\rightarrow\infty$ limit at fixed $k$ is cutoff by the logarithm.
This second limit is dual to the limit considered in Eq. \ref{eq:CSNtwo},
so this is the same finiteness we saw earlier, just obscured a bit by taking
the $1/N$ expansion.  Note also the $\lambda'\rightarrow 0,1$ limits are
$|k|\gg N$ and $|k|\ll N$ respectively,
so this is showing the $F\sim \min(k,N)^2\log(k+N)$ behavior pointed out above.

The upshot is that infinite series of pure CS theories are cutoff by
bounding the sphere free energy.  In other words, there are no infinite
series with bounded $F$.

\subsubsection*{$\mathcal{N}=2$ with $N_f$ multiplets }

We begin with the case of a single chiral superfield, i.e. $N_f=1$. This theory is a $\mathcal{N}=2$ superconformal theory without marginal deformations. There exist three supersymmetry breaking relevant deformations. In this case the theory exhibits a self duality under the map 
\begin{equation}
    U(N)_k \rightarrow U(k+\frac{1}{2}-N)_{-k}\;.
\end{equation}

This duality acts on the `t Hooft coupling $\lambda=\frac{N}{k}$ as $\lambda\rightarrow 1-\lambda$. 
Note that this does not preserve the domain $\lambda\in (0,\infty)$.  Equivalently, it can lead to negative $N$,
which does not make sense.  The resolution suggested in \cite{aharony_baryons_2016} is that the $\mathcal{N}=2$ theories
with $N>|k|$ do not exist (they break supersymmetry and presumably are not conformal).  Thus we will assume $\lambda\in(0,1)$.
This still allows an infinite series in $k$ but this takes us to the better controlled $\lambda\rightarrow 0$ limit.

The free energy was calculated by localization in Ref. \cite{Giombi:2017txg}, giving an exact expression 
in terms of an integral over $N$ eigenvalues.  This has a large N limit
\begin{align}
F_{\mathcal{N}=2} &= N^2 F_{CS}^{(0)} + N B_{susy} + \ldots \\
B_{susy}(\lambda) &= 
 \frac{i}{4\pi\lambda}\left(Li_2(-e^{i\pi \lambda})-Li_2(-e^{-i\pi \lambda})\right) \\
 &= \frac{\log 2}{2} - \frac{\pi^2\lambda^2}{48} + \ldots .
\end{align}

Now, as in Ref. \cite{Giombi:2017txg} we grant that this $F_{CS}$ is the one in Eq. \ref{eq:giombiN2F}, in which
case the finiteness is not manifest in the $\lambda\rightarrow 1$ limit.
However, the duality $\lambda\rightarrow 1-\lambda$ again implies that this
difficulty is only apparent.  Since the matter contribution to the free energy
is positive for $\lambda\le 1/2$, the number of these theories with a fixed bound
on $F$ is bounded above by the number of pure CS theories.

\subsubsection*{$\mathcal{N}=0$ }
The dualities in the non-supersymmetric cases are more complicated. The self dualities of the supersymmetric theories originate from a reshuffling of the bosonic and fermionic parts of the multiplets. In the non-supersymmetric cases this leads to a duality between fermionic and bosonic theories. Here we will focus on a specific pair of such theories, the critical boson theory (CB) and the massless (or ``regular'') fermion theory (RF). These have the actions
\begin{align*}
    S_{\mathrm CB}&=S_{CS}+\int {\mathrm d}^3 x\left(D_\mu\ov{\phi}D^\mu \phi+\frac{\lambda_4}{4N}(\ov{\phi}\phi)\right)\\
    S_{\mathrm RF}&=S_{CS}+\int {\mathrm d}^3 x \ov{\psi}\slashed{D}\psi
\end{align*}
The duality then reads 
\begin{equation}
    U(N)_{\mathrm{CB},k} \  \Leftrightarrow \ U(k)_{\mathrm{RF},-n+1/2} 
\end{equation}
which corresponds to $\lambda_{CB} = 1-\lambda_{RF}$.  These theories have no marginal deformations (so the additional coupling $\lambda_4$ is fixed). We can again write large $N$ expansions
\begin{align}
F_{CB} &= N^2 F_{CS}^{(0)} + N B_{CB} + \ldots \\
F_{RF} &= N^2 F_{CS}^{(0)} + N B_{RF} + \ldots 
\end{align}
Exact expressions for $B_{CB}$ and $B_{RF}$ are conjectured in Ref. \cite{Giombi:2017txg}.
From Eq. (3.11) there, while they both vanish as $\lambda\rightarrow 1$, they remain positive for $\lambda < 1$.
Figure \ref{fig:my_label} shows that they are positive away from this limit.

Now we can use the duality to argue that the complete set of these theories
is covered by the $\lambda\le 1/2$ regime of CB and the $\lambda\le 1/2$ of RF.
For both, the matter contribution to $F$ is positive, so the same arguments apply.
We conclude that these sets also do not lead to infinite series with bounded $F$.

In \cite{Jain_2013,aharony_flows_2018} a second dual pair of theories RB and CF is studied
and argued to also
have fixed points.  They furthermore argue that these fixed point theories can
flow down to the CB or RF theories with the same $(N,k)$.  By the F-theorem
\cite{jafferis_towards_2011}, they must have larger $F$ so they do not
contain infinite series with bounded $F$.

\section{Structures from free QFTs}
In this appendix we give another example of a structure arising from QFTs. We study free scalar fields in a $d$-dimensional Euclidean spacetime. As we are dealing with a free theory, the n-point correlators can be computed using Wick's theorem as a product over the two-point functions. The whole structure will thus be given by  sums of products of the two-point transition functions
\begin{equation}
    W^d(x,y)=\bra{x}e^{- H t}\ket{y}\,.
\end{equation}
We begin with the two-point transition function of a Klein-Gordon field of mass $m$ in flat space. For a $d$-dimensional theory it is given by \cite{Bertola:1999vq}
\begin{equation}
     W^d(x,y)=(2\pi)^{-d/2}\left(\frac{(x-y)^2}{\sqrt{m}}\right)^{\frac{d-2}{2}}K_{\frac{d-2}{2}}(\sqrt{m}(x-y)^2)\;.
\end{equation}
$K_a(x)$ denotes the modified Bessel function of the second kind. As already discussed in \S \ref{sec:QFTs_on_points} and part I, these functions are in the generic case exponential periods. The Conjecture 1, introduced in \S \ref{sec:QFTs_on_points}, implies that they are definable in some o-minimal structure. Only in the special cases of $d=1$ or $d=3$ the function reduces to
\begin{equation}
    K_{1/2}(x)=e^{-x}\sqrt{\pi/(2x)}\,,
\end{equation}
and definability in $\bbR_{\rm exp}$ is apparent.

For the Klein-Gordon field in AdS spacetime the two-point function was first computed in \cite{Fronsdal:1974ew} with the result
\begin{equation}
    W^d(x,y)=\frac{e^{-i\pi (d-1)/2}}{(2\pi)^{(d+1)/2}}((x\cdot y)^2-1)^{-\frac{d-1}{4}}Q^{\frac{d-1}{2}}_{\sqrt{d^2/4+m^2}-1/2}(x\cdot y).
\end{equation}
Here $Q_a^b(x)$ denotes the Legendre function of the second kind. Note that the signature only influences the possible signs of $x\cdot y$. The Legendre function can be expressed in terms of the hypergeometric functions as
\begin{equation}
    Q^\mu_\lambda(x)=\frac{\sqrt{\pi}\Gamma[\lambda+\mu+1]}{2^{\lambda+1}\Gamma[\lambda+\frac{3}{2}]}\frac{e^{i\pi \mu}(x^2-1)^{\mu/2}}{x^{\lambda+\mu+1}}\pFq{2}{F}{1}{\frac{\lambda+\mu+1}{2},\frac{\lambda+\mu+2}{2}}{\lambda+\frac{3}{2}}{\frac{1}{x^2}}
\end{equation}
The only dependence on the mass is in the lower argument of the Legendre function, $\lambda=\sqrt{d^2/4+m^2}-1/2$. The tameness in these parameters depends on the dimension and the signature of the distance $x\cdot y$. In even dimensions it is always non-tame, while in odd dimensions it is non-tame for  $x\cdot y>0$.

But the massive Klein-Gordon field with generic coupling is not conformal. The conformal limit is given by $\lambda\rightarrow 0$. In this case, the Legendre functions reduce to algebraic functions. The situation for positively curved spacetimes is similar, the only difference being the Legendre $Q$ function being replaced by a Legendre $P$ function.

\bibliography{literature}
\bibliographystyle{utphys}
\end{document}